\documentclass[a4paper,11pt]{article}
\pdfoutput=1

\usepackage{jcappub}
\usepackage{amsmath,amssymb,color,comment,epsfig,graphics,graphicx,lipsum,mathrsfs,xcolor}

\begin{document}

\title{
Gravitational effects on fluid dynamics in cosmological first-order phase transitions
}

\author[a]{Ryusuke Jinno}
\author[b,c,d]{and Jun'ya Kume}
\affiliation[a]{\small Department of Physics, Kobe University, Kobe 657-8501, Japan}
\affiliation[b]{\small
Dipartimento di Fisica e Astronomia ``G. Galilei'', Universit\`a degli Studi di Padova, via Marzolo 8, I-35131 Padova, Italy
}
\affiliation[c]{\small
INFN, Sezione di Padova, via Marzolo 8, I-35131 Padova, Italy
}
\affiliation[d]{\small
Research Center for the Early Universe (RESCEU), Graduate School of Science, The University of Tokyo, Tokyo 113-0033, Japan
}

\emailAdd{jinno@phys.sci.kobe-u.ac.jp}
\emailAdd{junya.kume@unipd.it}

\subheader{
{\rm RESCEU-12/24, KOBE-COSMO-24-02}
}

\abstract{
Cosmological first-order phase transition (FOPT) sources the stochastic gravitational wave background (SWGB) through bubble collisions, sound waves, and turbulence.
So far, most studies on the fluid profile of an expanding bubble are limited to transitions that complete in a much shorter time scale than the cosmic expansion.
In this study, we investigate gravitational effects on the fluid profile beyond the self-similar regime.
For this purpose we combine a hydrodynamic scheme in the presence of gravity with a fluid computation scheme under energy injection from the bubble wall. By performing (1+1)d simulations of spherical bubble for constant wall velocities, we find that the fluid generally develops a thinner shell in our cosmological setup, which qualitatively agrees with previous studies discussing the late-time behavior of fluid in expanding spacetime. 
We also observe reduction in the energy budget for the fluid kinetic energy. Furthermore, we find that the fluid profile develops sub-structure for accelerating bubble walls.
We also comment on the possible broadening of the SGWB spectral plateau.
}

\maketitle

\section{Introduction}
\setcounter{equation}{0}

Phase transitions are interesting phenomena widely observed in Nature, and they can also occur in the high-energy plasma in the early Universe.
First-order phase transition (FOPT) is among the most interesting of them, that involves supercooling and the release of latent heat.
Although the two phase transitions in the Standard Model (SM) of particle physics, the electroweak and quark-hadron transitions, have been found to be crossovers, the possibility of a FOPT in the early Universe~\cite{Hogan:1983ixn,Witten:1984rs,Hogan:1986qda} has been considered as a mechanism to solve mysteries whose solutions call for physics beyond the SM.

A FOPT in the early Universe first leads to the nucleation of tiny seeds of bubbles of the order parameter field due to thermal fluctuations or tunneling effects.
As soon as they nucleate, they start to expand and reach a cosmological size, driving the surrounding plasma in motion.
They eventually collide with each other and the transition completes, but the motion of the surrounding plasma remains in space even after that.
This process drives deviations from thermal equilibrium and other dynamics, and has in fact attracted much attention as a mechanism for the production of the baryon asymmetry of the Universe~\cite{Kuzmin:1985mm,Cohen:1993nk,Konstandin:2011ds,Servant:2014bla,Katz:2016adq,Cline:2020jre,Dorsch:2021ubz,Azatov:2021irb,Baldes:2021vyz}, dark matter~\cite{Witten:1984rs,Falkowski:2012fb,Hambye:2018qjv,Baldes:2020kam}, primordial black holes~\cite{Hawking:1982ga,Kodama:1982sf,Moss:1994iq,Khlopov:1998nm,Jedamzik:1999am}, primordial magnetic fields~\cite{Hogan:1983zz,Quashnock:1988vs,Vachaspati:1991nm}, topological defects~\cite{Kibble:1976sj,Kibble:1995aa,Borrill:1995gu}, and so on.

One of the distinguishing properties of FOPTs is their production of gravitational waves (GWs).
The aforementioned process of bubble nucleation, expansion, and collision as well as the plasma motion after that produces a sizable amount of GWs~\cite{Kosowsky:1991ua,Kosowsky:1992rz,Kosowsky:1992vn,Kamionkowski:1993fg,Huber:2008hg,Bodeker:2009qy,Jinno:2016vai,Jinno:2017fby,Konstandin:2017sat,Cutting:2018tjt,Cutting:2020nla}\cite{Hindmarsh:2013xza,Hindmarsh:2015qta,Hindmarsh:2017gnf,Cutting:2019zws,Jinno:2020eqg,Jinno:2022mie}\cite{Hindmarsh:2016lnk,Hindmarsh:2019phv,RoperPol:2023dzg}\cite{Brandenburg:2017rnt,Brandenburg:2017neh,Niksa:2018ofa,RoperPol:2019wvy}.
FOPTs in the TeV scale, where new physics beyond the SM may emerge, generally predict stochastic GW backgrounds (SGWBs) in the mHz-Hz band, see e.g. Refs.~\cite{Caprini:2015zlo,Weir:2017wfa,Mazumdar:2018dfl,Caprini:2018mtu,Caprini:2019egz,Caldwell:2022qsj,LISACosmologyWorkingGroup:2022jok}.
Space interferometers including LISA~\cite{2017arXiv170200786A}, DECIGO~\cite{Kawamura:2006up}, Taiji and TianQin, Einstein Telescope (ET)~\cite{Punturo:2010zz}, Cosmic Explorer (CE)~\cite{Reitze:2019iox}, as well as atomic interferometries have been proposed in this frequency range.
On the other hand, those around the QCD scale could source SGWB in the Pulsar Timing Array (PTA) band.
Recently, PTA collaborations (NANOGrav, EPTA/InPTA, PPTA and CPTA) have reported evidence for an isotropic SWGB as the expected Hellings-Downs (HD) angular correlation between pulsars’ line of sight were measured at $3-4 \sigma$ confidence level~\cite{NANOGrav:2023gor,EPTA:2023fyk,Reardon:2023gzh,Xu:2023wog}.
Interestingly, SGWB from FOPT can explain the current dataset relatively well if bubbles expand to nearly the scale of cosmological horizon (see e.g.~\cite{NANOGrav:2023hvm}).
However, the GW templates currently used in those analyses are only marginally valid as they are derived without taking the effect of gravity into account.

Bubbles of macroscopic size are in general subject to the effect of gravity.
Especially in transitions with an extreme amount of supercooling (e.g. Ref.~\cite{Konstandin:2011dr}), that are observationally interesting from the viewpoint of GW production, the effect of gravity can be non-negligible because the bubbles tend to expand to a comparable size to the Hubble radius\footnote{
In such extremely strong transitions, the effect of particle splitting can be crucial for the bubble to develop fluid profile~\cite{Bodeker:2017cim,Azatov:2020ufh,Gouttenoire:2021kjv,Azatov:2023xem} rather than to run away~\cite{Bodeker:2009qy}.
}.
For a spherically symmetric bubble expanding in the absence of gravity, self-similar solution is enough to estimate the fluid profile and the energy budget, because the only macroscopic scale in the system is the size of the bubble (see e.g. Refs.~\cite{Espinosa:2010hh,Ellis:2019oqb,Giese:2020znk,Giese:2020rtr}).
Along this line, the effect of cosmological expansion on the self-similar fluid profile was investigated in Ref.~\cite{Cai:2018teh} by taking late-time limit of the general relativistic equation of motion. More recently, fluid profile around a single bubble of cosmological size was investigated with full general relativity, under the assumption of self-similarity~\cite{Giombi:2023jqq}.
To be precise, however, the system does not enjoy self-similarity as a new scale -- the Hubble scale -- comes into play.
Therefore, the aim of the present paper is to investigate the effect of gravity on these bubbles without assuming self-similarity and to obtain implications to the SGWB spectrum from sound waves.
Studies in this direction are important in order to finally extract the information on the underlying particle physics from the GW signals once observed~\cite{Hashino:2018wee,Gowling:2021gcy,Gowling:2022pzb,Caprini:2024hue}.

For this purpose, we embed hydrodynamic simulation for FOPTs into the cosmological hydrodynamics scheme of Ref.~\cite{Noh:2018sil}.
For the former we adopt (a slightly modified version of) the Higgsless scheme proposed in Ref.~\cite{Jinno:2020eqg,Jinno:2022mie}.
The central idea of this scheme is to ``integrate out'' the dynamical scalar field driving the transition and reduce the relevant dynamics into that of the fluid alone reacting to the energy injection at moving boundaries.
This scheme is thus free from microscopic scales and is able to simulate the system with lower computational cost.
On the other hand, the latter is to solve the peculiar motion of fluid around the Friedmann-Lema\^itre-Robertson-Walker (FLRW) spacetime with perturbatively taking into account the fluid backreaction to the spacetime.
Therefore, combining these two schemes allows us to evolve the fluid and bubble for long time enough to see gravitational effects on fluid dynamics in cosmological FOPT.

The rest of the paper is organized as follows. 
In Sec.~\ref{sec:Higgsless}, we give a brief review of the Higgsless scheme for the (1+1)d hydrodynamic simulation in a flat spacetime background.
In Sec.~\ref{sec:with_GR}, we introduce the cosmological hydrodynamics and then combine these two schemes to solve the fluid dynamics in FOPTs without assuming self-similarity.
Then in Sec.~\ref{sec:results}, we present the numerical results obtained and compare them with the known solutions in Sec.~\ref{sec:comparison}. 
Finally in Sec.~\ref{sec:dc} we discuss implications to GW signals and then conclude.

\section{Hydrodynamic scheme for FOPTs}
\setcounter{equation}{0}
\label{sec:Higgsless}

In this section, we briefly explain how we perform (1+1)d hydrodynamic simulations with energy injection at the moving boundary.
We follow the idea of the Higgsless scheme proposed in Refs.~\cite{Jinno:2020eqg,Jinno:2022mie}, with a slight modification in the choice of fundamental variables.
This approach allows us to study the fluid dynamics around the true-vacuum bubble without introducing the dynamical Higgs field.
Then we briefly describe the well-known self-similar solutions obtained in this scheme.

\subsection{Rewriting the energy-momentum conservation}

The starting equation is the energy-momentum conservation for the fluid
\begin{align}
\nabla_\mu T^{\mu \nu} 
&= 0,
\label{eq:EMconservation}
\end{align}
with the energy-momentum tensor for perfect fluid
\begin{align}
T^{\mu \nu}
&= 
w u^\mu u^\nu + p g^{\mu \nu},
\qquad
w
=
\rho + p.
\end{align}
Throughout the paper we use the metric convention $g_{\mu \nu} = {\rm diag} (- 1, + 1, + 1, + 1)$ and set the speed of light to unity as $c = 1$.
The fluid velocity is $u^\mu = \gamma (1, v^i)$ with $\gamma = 1 / \sqrt{1 - v^2}$ being the relativistic factor.
We take the bag equation of state
\begin{align}
\rho
&=
a T^4 + \epsilon,
\qquad
p
=
\frac{a}{3} T^4 - \epsilon,
\qquad
w
=
\frac{4}{3} a T^4,
\end{align}
with $a$ being constant.
The vacuum energy takes $\epsilon > 0$ ($\epsilon = 0$) in the symmetric (broken) phase.
Moving to $(t, r, \theta, \phi)$ coordinate, Eq.~(\ref{eq:EMconservation}) reduces to
\begin{align}
\partial_t (w \gamma^2  - p) + \partial_r (w \gamma^2 v) + \frac{2}{r} (w \gamma^2 v)
&=
0,
\\
\partial_t (w \gamma^2 v) + \partial_r (w \gamma^2 v^2 + p) + \frac{2}{r} (w \gamma^2 v^2)
&=
0.
\end{align}
Here $v \equiv v^r$ is the radial velocity, as we consider a spherical bubble.
The question is which variables we should take as fundamental ones.
For example, if we choose $(w, v)$, the relation between $(w, v)$ and the quantities in the time derivative, $w \gamma^2 - p$ and $w \gamma^2 v$, change discontinuously at the wall position.
Indeed, the relation $p = w / 4 - \epsilon$ changes across the wall due to the change in $\epsilon$.
Such a setup is difficult to implement numerically.

In the Higgsless scheme, the dynamics of the Higgs field is integrated out and incorporated into the system as one of the boundary conditions~\cite{Jinno:2020eqg,Jinno:2022mie}, under the assumption that the wall reaches a terminal velocity as soon as bubbles grow to a cosmological size.
Here we use a slightly different choice for the fundamental variables from these references.
Since the vacuum energy behaves as
\begin{align}
\epsilon
&=
\epsilon_0 \Theta (r - r_w (t)),
\end{align}
the time derivative of the vacuum energy can be traded off with its spatial derivative as
\begin{align}
\partial_t \epsilon
&=
- \dot{r}_w \partial_r \epsilon.
\end{align}
Thus the energy-momentum conservation can be rewritten as
\begin{align}
\partial_t \left( w \gamma^2 - \frac{w}{4} \right) + \partial_r (w \gamma^2 v - v_w \epsilon) + \frac{2}{r} (w \gamma^2 v)
&=
0,
\label{eq:deq_flat1}
\\
\partial_t (w \gamma^2 v) + \partial_r \left( w \gamma^2 v^2 + \frac{w}{4} - \epsilon \right) + \frac{2}{r} (w \gamma^2 v^2)
&=
0.
\label{eq:deq_flat2}
\end{align}
Though discontinuities appear in the flux, they are easy to implement with the Kurganov-Tadmor scheme as we explain below.
Therefore, we solve these differential equations in terms of the fluid energy-momentum tensor
\begin{align}
T^t
&\equiv
T^{tt}
=
w \gamma^2 - \frac{w}{4},
\qquad
T^r
\equiv
T^{tr}
=
w \gamma^2 v.
\end{align}
It is straightforward to express $w$ and $v$ in terms of these quantities
\begin{align}
w
&= \frac{4}{3}\left(\sqrt{4(T^t)^2 - 3(T^r)^2} -T^t\right),
\qquad
v
= \frac{2 T^t - \sqrt{4 (T^t)^2 - 3 (T^r)^2}}{T^r}.
\end{align}
As a result, we get the following equations of motion for the fundamental variables $(T^t, T^r)$
\begin{align}
&\partial_t T^t
+
\partial_r
\left(
T^r - v_w \epsilon
\right)
+
\frac{2}{r} T^r
=
0,
\label{eq:deq_flat_T1_bare}
\\
&\partial_t T^r
+
\partial_r \left(
\frac{5}{3} T^t - \frac{2}{3} \sqrt{4 (T^t)^2 - 3 (T^r)^2}
-
\epsilon
\right)
+
\frac{2}{r}
\left(
2 T^t - \sqrt{4 (T^t)^2 - 3 (T^r)^2}
\right)
=
0.
\label{eq:deq_flat_T2_bare}
\end{align}
For later purpose it is convenient to define
\begin{align}
\tilde{T}^t
&\equiv
\frac{T^t}{w_\infty},
\qquad
\tilde{T}^r
\equiv
\frac{T^r}{w_\infty}.
\end{align}
The corresponding differential equations are
\begin{align}
&\partial_t \tilde{T}^t
+
\partial_r
\left(
\tilde{T}^r - \frac{3}{4} v_w \alpha
\right)
+
\frac{2}{r} \tilde{T}^r
=
0,
\label{eq:deq_flat_T1}
\\
&\partial_t \tilde{T}^r
+
\partial_r \left(
\frac{5}{3} \tilde{T}^t - \frac{2}{3} \sqrt{4 (\tilde{T}^t)^2 - 3 (\tilde{T}^r)^2}
-
\frac{3}{4} \alpha
\right)
+
\frac{2}{r}
\left(
2 \tilde{T}^t - \sqrt{4 (\tilde{T}^t)^2 - 3 (\tilde{T}^r)^2}
\right)
=
0,
\label{eq:deq_flat_T2}
\end{align}
where we expressed the rescaled vacuum energy $\tilde{\epsilon} \equiv \epsilon / w_\infty$ with the $\alpha$ parameter
\begin{align}
\alpha
&\equiv
\frac{4}{3} \frac{\epsilon}{w_\infty}
=
\frac{4}{3} \tilde{\epsilon}.
\end{align}
Note that our $\alpha$ parameter is defined both in the symmetric and broken phases, as $\epsilon$ is defined in both phases.
The $\alpha$ parameter in the literature (e.g. Ref.~\cite{Espinosa:2010hh}) corresponds to our $\alpha$ in the symmetric phase.\footnote{
See e.g. Refs.~\cite{Hindmarsh:2015qta,Hindmarsh:2017gnf,Hindmarsh:2019phv,Giese:2020rtr,Giese:2020znk} for other choices to characterize the vacuum energy.
}

In the end, these equations of motion for $(\tilde{T}^t, \tilde{T}^r)$ are solved with the pre-defined value of $\alpha$ in simulating (fast) FOPTs. As mentioned above, this $\alpha$ admits the spatial discontinuity such as $\alpha > 0$ or $\alpha = 0$, depending on $r > r_w (t)$ or $r < r_w (t)$. Therefore, one needs to implement a shock capturing scheme in order to solve this system.

\subsection{Taking care of discontinuities}

In this study, we use the Kurganov-Tadmor scheme~\cite{KURGANOV2000241} to take care of the discontinuity around the bubble wall. 
This scheme has been used to investigate fluid dynamics in FOPTs in e.g. Refs.~\cite{Jinno:2019jhi,Jinno:2020eqg,Jinno:2022mie}.
Following Ref.~\cite{Jinno:2020eqg}, here we briefly describe this method.
Our differential equations~\eqref{eq:deq_flat_T1} and~\eqref{eq:deq_flat_T2} have the form
\begin{align}
\partial_t u + \partial_r f + g
&=
0,
\end{align}
with $f$ being dependent on $u$, and $g$ dependent on $u$ and $r$.
Let $u_j^n$ denote the $u$ variable at the $n$-th temporal grid point and the $j$-th spatial grid point.
In this scheme, the Euler step is calculated as
\begin{align}
u_j^{n + 1}
&=
u_j^n
-
\frac{\Delta t}{2 \Delta r} \left[ H_{j + 1 / 2}^n - H_{j - 1 / 2}^n \right]
-
g (u_j^n, r^j).
\end{align}
The flux $H$ is calculated from $u$ defined at the half grid $j \pm 1 / 2$ through
\begin{align}
H_{j + 1 / 2}
&=
\frac{f (u_{j + 1 / 2}^+) + f (u_{j + 1 / 2}^-)}{2}
-
\frac{a_{j + 1 / 2}}{2} \left[
u^+_{j + 1 / 2} - u^-_{j + 1 / 2}
\right].
\end{align}
The staggered values $u^\pm_{j + 1 / 2}$ are estimated from the integer grid as
\begin{align}
u^+_{j + 1 / 2}
&=
u_{j + 1}
-
\frac{\Delta r}{2} (u_r)_{j + 1},
\qquad
u^-_{j + 1 / 2}
=
u_j
+
\frac{\Delta r}{2} (u_r)_j.
\end{align}
The maximal velocity $a_{j + 1 / 2}$ should in principle be the largest absolute eigenvalue of the matrix $\partial f / \partial u$ evaluated at $u = u^\pm_{j + 1 / 2}$, but for practical purpose we set $a_{j + 1 / 2} = c_s = 1 / \sqrt{3}$.
The spatial derivative $u_r$ at the integer grid $(u_r)_j$ is estimated with the nonlinear minmod function
\begin{align}
(u_r)_j
&=
{\rm minmod}
\left(
\theta \frac{u_j - u_{j - 1}}{\Delta r},
\frac{u_{j + 1} - u_{j - 1}}{2 \Delta r},
\theta \frac{u_{j + 1} - u_j}{\Delta r}
\right).
\end{align}
This function returns a nonzero value only when the three arguments have the same sign
\begin{align}
{\rm minmod} (a, b, c)
&=
{\rm (sign~of~the~three~elements)}
\times
{\rm min} (|a|, |b|, |c|).
\end{align}
The hyperparameter $\theta$ is chosen arbitrarily from $1 \leq \theta \leq 2$ and we set $\theta = 2$ throughout our simulation.\footnote{
Although not shown below, we confirm that our results have negligible dependence on $\theta$.
}
The Euler step described above can be promoted to the commonly used 4th order Runge-Kutta method (see {\it e.g.} Ref.~\cite{Jinno:2020eqg,Jinno:2022mie} for more details), which is also adopted in the following numerical analysis.

\subsection{Numerical results for the flat spacetime}\label{sec:results_flat}

\begin{figure}[htbp]
\begin{center}
\includegraphics[width=\columnwidth]{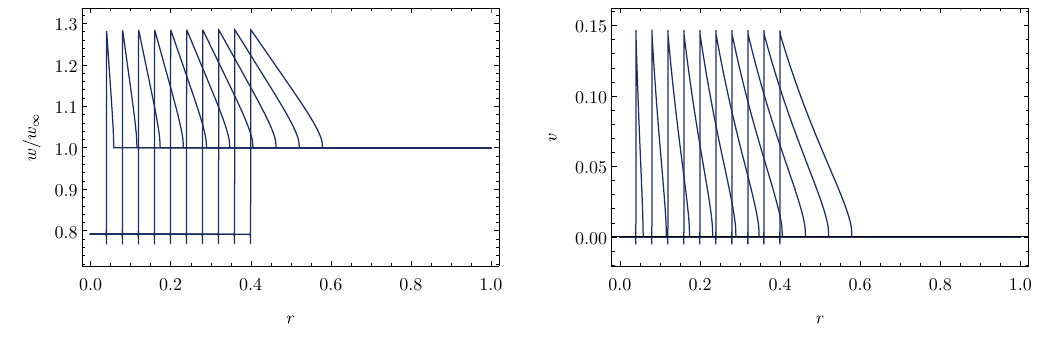}
\includegraphics[width=\columnwidth]{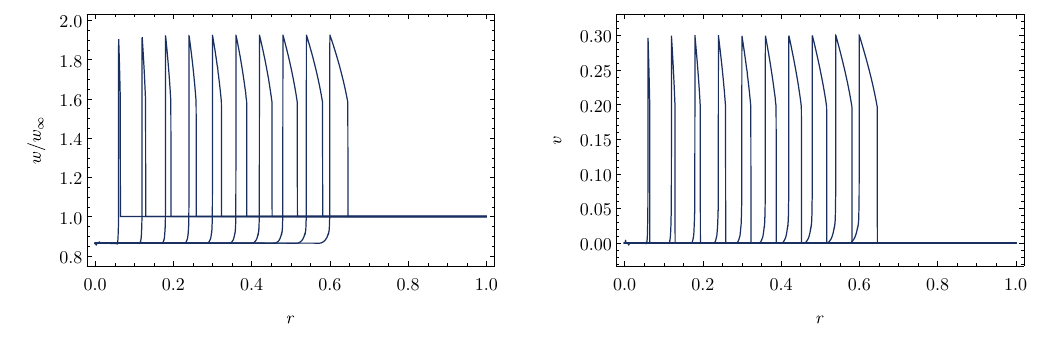}
\includegraphics[width=\columnwidth]{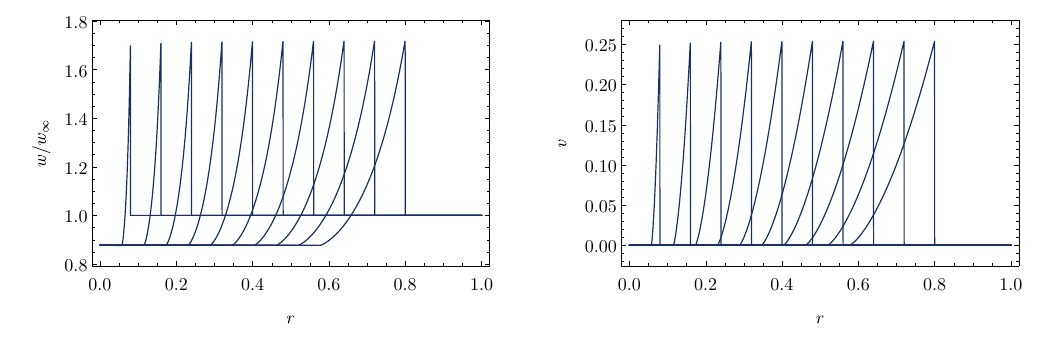}
\caption{\small
Fluid profile simulated for $v_w = 0.4$, $0.6$, and $0.8$ from top to bottom with $\alpha = 0.1$.
The left column is $\tilde{w} = w / w_\infty$, while the right is the outward velocity $v$.
We take $n_r = 10000$ spatial grid points and evolve the system for $n_t = 20000$ time steps with $[\Delta r]_{\rm sim} = 10^{- 4}$ and $[\Delta t]_{\rm sim} = 0.5 [\Delta r]_{\rm sim}$.
}
\label{fig:flat}
\end{center}
\end{figure}

We show in Fig.~\ref{fig:flat} the time evolution of a spherical expanding bubble for $v_w = 0.4$, $0.6$, and $0.8$ with $\alpha = 0.1$, which are obtained in the numerical scheme described above.
We take $n_r = 10000$ grid points and evolve the system for $n_t = 20000$ time steps with $[\Delta r]_{\rm sim} = 10^{-4}$ and $[\Delta t]_{\rm sim} = 0.5[\Delta r]_{\rm sim}$.
The radial coordinate at spatial grid $j$ is set to be $r_j = (j + 0.5) \Delta r$ (with $j = 0, 1, \cdots$), and the boundary condition is Neumann $\partial_r \tilde{K}^t = \partial_r \tilde{K}^r = 0$ both at $r = 0$ and $r = \infty$.
Starting from the top panel, we can see that the fluid forms three different profiles called deflagration, hybrid, and detonation. Note that the profile developed depends on the values of $\alpha$ and $v_w$.
For more details on these profiles, see Ref.~\cite{Espinosa:2010hh} for example.
By scaling the spatial coordinate as $r/t$, one can explicitly see the self-similarity in all the profiles.

This self-similarity is, however, manifested by ignoring the scale of the expansion rate of the universe and working in the flat spacetime. 
Such a flat spacetime approximation would be appropriate if the transition completes very quickly compared to the expansion of the universe.
This is not the case for some models that predict extremely supercooled phase transitions that would complete on a time scale comparable to the Hubble time (e.g. Ref.~\cite{Konstandin:2011dr}).
In order to evaluate the dynamics of fluid (and its contribution to the SGWB) in such situation, one needs to solve the system general relativistically and to go beyond the self-similar approximation. 

\section{Fluid dynamics in FOPTs beyond the self-similarity}
\setcounter{equation}{0}
\label{sec:with_GR}

In this section, we introduce the cosmological hydrodynamic scheme proposed in Ref.~\cite{Noh:2018sil}, which solves the peculiar motion of fluid with relativistic pressure 
in a (nearly) Friedmann-Lema\^itre-Robertson-Walker (FLRW) spacetime.
Then we explain how to incorporate the Higssless scheme into this scheme to investigate the fluid dynamics around the slowly expanding bubble wall.

\subsection{Hydrodynamic equations around the FLRW spacetime}\label{sec:cosmohydro}

We assume the following spacetime metric following Ref.~\cite{Noh:2018sil}, in which the peculiar motion of relativistic fluid around the FLRW spacetime is considered:
\begin{align}
ds^2
&=
- (1 + 2 \Phi(t,r)) dt^2 - 2 \partial_r \chi(t, r) dt dr + (1 - 2 \Psi(t , r) )a(t)^2(dr^2 + r^2d\Omega^2).\label{eq:metric_pert}
\end{align}
Since we are interested in the fluid dynamics around a single bubble, spherical symmetry is assumed here.
After taking the uniform-expansion gauge (UEG) that assumes vanishing perturbed part of the trace of extrinsic curvature, the equations of motion for the fluid with relativistic pressure and metric perturbations are derived in Ref.~\cite{Noh:2018sil}. 

With the fluid four-velocity $u^{\mu} = \gamma(1, v(t,r)/a, 0, 0)$, we find the following equations of motion:
\begin{align}
\partial_t (w \gamma^2  - p) + \frac{1}{a}\left(\partial_r + \frac{2}{r}\right) (w \gamma^2 v) + (3 +v^2)Hw\gamma^2 + w \gamma^2 v \frac{1}{a}\partial_r \Phi
&=
0,\label{eq:GRPN_EOM1}
\\
\partial_t (w \gamma^2 v) + \frac{1}{a} \left\{ \partial_r (w \gamma^2 v^2 + p) + \frac{2}{r} (w \gamma^2 v^2) \right\} + 4Hw\gamma^2v + w \gamma^2 \frac{1}{a}\partial_r \Phi
&=
0,\label{eq:GRPN_EOM2}
\\
\frac{1}{a^2}\Delta\Phi -3\frac{\ddot{a}}{a} - \frac{1}{2 M_{\rm Pl}^2} (w + 2 p) - \frac{1}{M_{\rm Pl}^2} w \gamma^2 v^2
&=
0,\label{eq:GRPN_EOM3}
\end{align}
where we omit the EoM of $\Psi$ and the constraint equation of $\chi$ for simplicity.
These equations, linearized in terms of the metric perturbations, are derived under the following assumptions:
\begin{equation}
    \Phi \ll 1, \quad \Psi \ll 1, \quad \gamma^2 l_c^2 H^2 \ll 1, \label{eq:conditions}
\end{equation}
where $l_c$ is the characteristic length scale of the dynamics.
The first two conditions are for the smallness of the gravitational backreaction, while the last condition can be referred to as the short-scale condition.

Following Ref.~\cite{Noh:2018sil}, we perturbatively decompose the fluid quantities into background and perturbation as
\begin{align}
  \rho(t, r) &= \rho_b(t) + \delta\rho(t, r),\\
  p(t, r) &= p_b(t) + \delta p(t, r),\\
  w(t, r) &= w_b(t) + \delta w(t, r),
\end{align}
where the homogeneous background parts accounts for the evolution of the scale factor $a(t)$ in the above equations.
Then, as we will see below, one can derive the equations of motion for the perturbations including the gravitational potential $\Phi$.
In this way, this formalism allows us to solve the peculiar motion of relativistic fluid that constitutes the FLRW spacetime, while perturbatively taking into account the gravitational backreaction.

Before proceeding, we comment on the gauge choice in the present paper.
As mentioned in the beginning of this section, we follow Ref.~\cite{Noh:2018sil} and adopt the UEG, in which the trace of the extrinsic curvature $\kappa$ is set to zero.
This choice, together with the spatial gauge condition, fixes the gauge completely and leaves no further gauge degrees of freedom. Consequently, all the remaining variables can be regarded as gauge invariant quantities, ensuring the absence of gauge artifacts~\cite{Noh:2004bc, Hwang:2012aa}. 
Notice that the UEG is not a common choice in the context of cosmological perturbation theory. For interested readers, we provide a brief discussion in App.~\ref{sec:gauge_choice}, where we compare physical quantities in the UEG and in the zero-shear gauge (ZSG), which is more standard in this context. In the regime of interest, the differences between the quantities in UEG and ZSG are expected to be suppressed by a common factor related to the thickness of fluid shell. 

Despite the absence of gauge artifacts, one may question whether the quantities presented in this paper ({\it e.g.}, fluid velocity and enthalpy) are suitable for direct comparison with their counterparts in flat spacetime. That is, it might be possible that quantities in one gauge is more suited than another for comparison with quantities in the flat background.
This concern reflects a broader issue in general relativity, where the meaning of physical quantities is inherently tied to the gauge choice.
In this regard, we note that the UEG is the one that makes the expansion rate uniform over each spatial slice. This property isolates local fluid motions from spatial variations in the expansion rate that would be present in other gauges, providing a clearer interpretation of fluid dynamics in an expanding spacetime.
Thus, we conclude that the present gauge choice is at least well-suited, albeit not proven to be the best, for comparisons with results in flat spacetime.

In the following, we apply this scheme to the cosmological FOPTs in order to solve the fluid dynamics in response to the energy injection at the phase front. It should be noted that the validity of the calculation is restricted to the region that satisfies the conditions in Eq.~\eqref{eq:conditions}.
We will clarify in what situation these conditions are satisfied.

\subsection{Application to FOPTs}
\label{sec:cosmo_hydro_FOPT}

Here we describe how to embed the Higgsless scheme into the cosmological hydrodynamics scheme to solve the cosmological fluid dynamics around the bubble. 
Let us assume that the background part consists of the radiation and the vacuum energy, such as the outside of bubble wall. The background quantities are respectively given as 
\begin{equation}
    \rho_b = \frac{3}{4}w_b + \epsilon, \quad p_b = \frac{1}{4}w_b - \epsilon, \quad w_b \propto a(t)^{-4}.
\end{equation}
In this case, the background scale factor $a(t)$ shows a relatively complicated behavior.
The Friedmann equation
\begin{align}
H^2
&=
H_0^2
\left[
\frac{1}{1 + \alpha_0} \left( \frac{a}{a_0} \right)^{- 4}
+
\frac{\alpha_0}{1 + \alpha_0}
\right],
\end{align}
has an analytical expression
\begin{align}
\frac{a(t)}{a_0}
&=
\alpha_0^{-\frac{1}{4}}
\sqrt{\sinh\left( 2H_0t\sqrt{\frac{\alpha_0}{1 + \alpha_0}} + {\rm arctanh}\left( \sqrt{\frac{\alpha_0}{1 + \alpha_0}}\right)\right)},\label{eq:scale_factor_RV}
\end{align}
where $H_0$ is the initial Hubble parameter at $t = 0$ and $\alpha_0 = \epsilon/\rho_{\rm rad}(t = 0)$. Throughout this paper, we assume that the true-vacuum bubble nucleates during the radiation domination ($\alpha_0 \ll 1$), which is actually important for satisfying the weak backreaction condition.
In App.~\ref{sec:tv_as_bg}, we summarize the alternative formulation and the results for the case where only radiation is taken as background. Though there exists a certain complication in this decomposition compared to the above one, we found that the different background settings do not change our main results.

In this scheme, the dynamics caused by the energy released at the bubble wall can be described as the perturbation with respect to this background.
By introducing $\alpha(t) = \alpha_0 (a(t) / a_0)^4$, the perturbative quantities can be expressed as 
\begin{align}
\delta \tilde{p} = \frac{1}{4}\delta \tilde{w} + \frac{3}{4}\alpha(t)\Theta(r_w(t) -r),
\\
\delta \tilde{\rho} = \frac{3}{4}\delta \tilde{w} - \frac{3}{4}\alpha(t)\Theta(r_w(t) -r),
\end{align}
where $\tilde{X}$ indicates that $X$ is normalized by the background enthalpy as $\tilde{X} = X/w_b$, and the last terms are introduced so that the vacuum energy contribution vanishes inside the wall ($r < r_w(t)$).
Here in the expanding universe, the wall position is evaluated as
\begin{align}
r_w(t)
&=
\int dt~\frac{v_w}{a(t)} = \int d \eta~v_w,
\label{eq:wall_position}
\end{align}
where the co-moving time $\eta$ is introduced for the latter convenience and the metric perturbation is ignored since it only gives sub-leading effects.

With the above background dynamics and the expressions for the perturbative quantities, the EoMs~\eqref{eq:GRPN_EOM1}--~\eqref{eq:GRPN_EOM3} are expanded to the linear order in $\delta \tilde{w}$ and $v$. To solve them, however, we need to take care of discontinuities arising from the moving wall. We found that the method in Sec.~\ref{sec:Higgsless} works in the same way.
In analogy with those defined in Sec.~\ref{sec:Higgsless}, let us introduce the auxiliary variables defined as
\begin{align}
\tilde{T}^t
&\equiv
\tilde{w} \gamma^2 - \frac{\tilde{w}}{4}
\qquad
\tilde{T}^r
\equiv
\tilde{w} \gamma^2 v,
\end{align}
which are related to $(\delta\tilde{w}, v)$ as
\begin{align}
\tilde{w} = 1 + \delta\tilde{w}
&= \frac{4}{3}\left(\sqrt{4(\tilde{T}^t)^2 - 3(\tilde{T}^r)^2} -\tilde{T}^t\right),
\qquad
v
= \frac{2 \tilde{T}^t - \sqrt{4 (\tilde{T}^t)^2 - 3 (\tilde{T}^r)^2}}{\tilde{T}^r}.
\end{align}
With this convention, we find the following equations of motion for $\tilde{T}^t$ and $\tilde{T}^r$ at the leading order of perturbation:
\begin{align}
&\partial_t \tilde{T}^t
+
\frac{1}{a (t)}
\left[
\left( \partial_r + \frac{2}{r} \right) (\tilde{w} \gamma^2 v)
-
\partial_r
\left(
\frac{3}{4}
\alpha(t) v_w(t)
\Theta(r - r_w(t))
\right)
+
\tilde{w} \gamma^2 v \, \partial_r \Phi
\right]
=
0,
\label{eq:eom1}
\\
&\partial_t \tilde{T}^r
+
\frac{1}{a (t)}
\left[
\left( \partial_r + \frac{2}{r} \right) (\tilde{w} \gamma^2 v^2)
+
\partial_r \left( \frac{\tilde{w}}{4} - \frac{3}{4} \alpha(t) \Theta(r - r_w(t)) \right)
+
\tilde{w} \gamma^2 \, \partial_r \Phi
\right]
=
0,
\label{eq:eom2}
\\
&\partial_r \Phi
=
\frac{w_b (t)}{M_{\rm Pl}^2}
\frac{a^2 (t)}{r^2}
\int_0^r dr'~r'^2
\left(
\frac{3}{4} (\tilde{w} - 1)
-
\frac{3}{4}
\alpha (t) (\Theta(r^{\prime} - r_w(t)) - 1)
+
\tilde{w} \gamma^2 v^2
\right).
\label{eq:eom_grav_pot}
\end{align}
The first and second equations give the response of the cosmic fluid in the expanding spacetime to an injection of energy from a moving wall. The wall contribution is again expressed by the Heaviside theta function $\Theta$, whose discontinuity can be dealt with the same way as we discussed in Sec.~\ref{sec:Higgsless}.
Most importantly, the strength of injection $\alpha(t)$ is now time-dependent and grows as $\propto a(t)^4$. Here the cosmic expansion rate matters and it is expected to give the qualitative difference from the flat spacetime solutions. 
On the other hand, the last equation \eqref{eq:eom_grav_pot} describes the profile of the gravitational potential $\Phi$.
Among the three terms in the parenthesis, the second term is the contribution from the vacuum energy that changes discretely across the bubble wall, while the first and third terms are from the fluid thermal/kinetic energy driven by the latent heat release.
In the language of cosmological FOPTs, the second term is of the order $\sim \alpha$ while the first and third are of the order $\sim \kappa \alpha$, where $\kappa$ is the kinetic energy fraction defined below (see Eq.~\eqref{eq:kinetic}).

In the following, these equations are numerically solved for $(\tilde{T}^t, \tilde{T}^r)$ and then the solutions are converted into $(\tilde{w}, v)$.
However, before moving on, let us give comments on the regime of validity of our analysis by referring to the conditions in Eq.~\eqref{eq:conditions}. 
As mentioned above, $\Phi$ (and in fact $\Psi$) gets larger for larger latent heat $\alpha(t)$.
The integration of the dominant second term in Eq.~(\ref{eq:eom_grav_pot}) can be performed as
\begin{align}
\partial_r \Phi
&\simeq
\frac{3}{4 M_{\rm Pl}^2} w_b (t) \alpha (t)
\frac{a^2 (t)}{r^2}
\int_0^r dr'~r'^2 (1 - \Theta(r^{\prime} - r_w(t)))
\nonumber \\
&=
\displaystyle\frac{1}{4 M_{\rm Pl}^2} w_b(0) \alpha_0
\frac{a^2 (t)}{r^2} \times 
\left\{
\begin{array}{cc}
r^3
&(r < r_w),
\\
r_w^3
&(r > r_w),
\end{array}
\right.
\end{align}
thus giving
\begin{align}
\Delta\Phi(t)
&\equiv
|\Phi(t, 0) - \Phi(t, r\to\infty)|
\simeq
\frac{3}{8 M_{\rm Pl}^2} w_b(0) \alpha_0 a^2(t)r^2_w(t).
\label{eq:grav_pot_bubble}
\end{align}
Therefore, in order for bubbles to expand over a cosmological time scale while satisfying $\Phi, \Psi \ll 1$, initial radiation dominance $\alpha_0 \ll 1$ is an important assumption.
Meanwhile, $\Delta\Phi(t) \equiv |\Phi(t,0) - \Phi(t, r\to\infty)|$ becomes sufficiently large if we run the simulation for longer time. Therefore, $\Delta\Phi(t) \lesssim {\cal O} (0.1)$ set an upper limit on the simulation time for which the results obtained are considered reasonable.

One also needs to care about the short-scale condition.
The last condition in Eq.~(\ref{eq:conditions}) is actually connected to the perturbativity of $\chi (t, r)$~\cite{Noh:2018sil}
\begin{align}
\frac{\partial_r \chi}{a}
&\ll
v,
\label{eq:chi_perturbative}
\end{align}
which is derived with the ADM momentum constraint equation $\partial_r^3 \chi / a^3 \sim w \gamma^2 v / M_{\rm pl}^2$.
This implies that the physical length scale $l_c$ should be estimated from the region where the fluid velocity $v$ is nonzero.
As long as the perturbativity condition for $\Phi$ is satisfied, it is safely taken to be the order of the shell thickness in the flat background, which is much smaller than the bubble size and the Hubble horizon.
However, once $\Phi$ becomes of ${\cal O} (0.1)$, fluid velocity starts to be induced over the Hubble size, and the system size can no longer be regarded as the shell thickness.
Since this transition is triggered by $\Phi \sim {\cal O} (0.1)$, we can monitor the violation of the short scale condition by keeping track of the perturbativity of $\Phi$. 

In addition to the above two conditions, too large $\alpha (t)$ for a fixed value of $v_w$ can also lead to unphysical solution $v > v_w$~\cite{Espinosa:2010hh}, and in some cases this appears before the perturbativity breakdown of $\Phi$.
From these considerations, we simulate the system until $t = t_{\rm max}$, when either the perturbativity condition of $\Phi$ or the physical condition on $\alpha (t)$ breaks down.
To go beyond and investigate the potential effects of strong gravitational fields, one needs to work with the full general relativity as in Refs.~\cite{Johnson:2011wt,Giombi:2023jqq}.

\section{Numerical results}\label{sec:results}

In this section, we show the time evolution of an expanding bubble with gravity for several scenarios, and discuss their characteristic features. 
We first consider constant wall velocity and present the numerical solutions as well as the efficiency factor for different values of the wall velocity and the radiation energy fraction at the time of bubble nucleation.
We then include the case with luminal wall velocity, since in extremely supercooled scenarios the bubbles reach almost the speed of light.
We finally consider time-dependent wall velocity.
The last scenario takes account of the decrease of the fluid pressure from the outside of the wall as the temperature drops with the expansion of the universe.
Throughout this section, we use geometric unit $M_{\rm Pl} = 1$.

\subsection{Constant-velocity walls}\label{sec:const}

In this subsection we assume the constant wall velocity $v_w = const.$ as in Sec.~\ref{sec:Higgsless}.
We first highlight the early-time solutions for $0 \leq t \leq H_0^{-1}$ to see the qualitative behavior of the system for sufficiently small $\Phi$.
Even at these early times, a significant difference from the flat spacetime solutions arises due to the cosmological expansion.
Then we perform simulations for a longer period to see the asymptotic behavior of the system until $\alpha (t)$ becomes of order unity, in which case GW production will be observationally relevant.

\subsubsection{Early time solutions}~\label{sec:comparison_flat}
\begin{figure}[htbp]
\begin{center}
\includegraphics[width=\columnwidth]{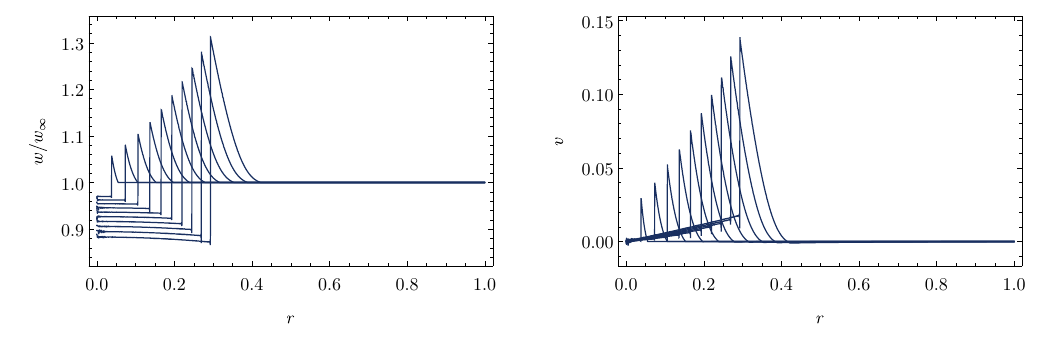}
\includegraphics[width=\columnwidth]{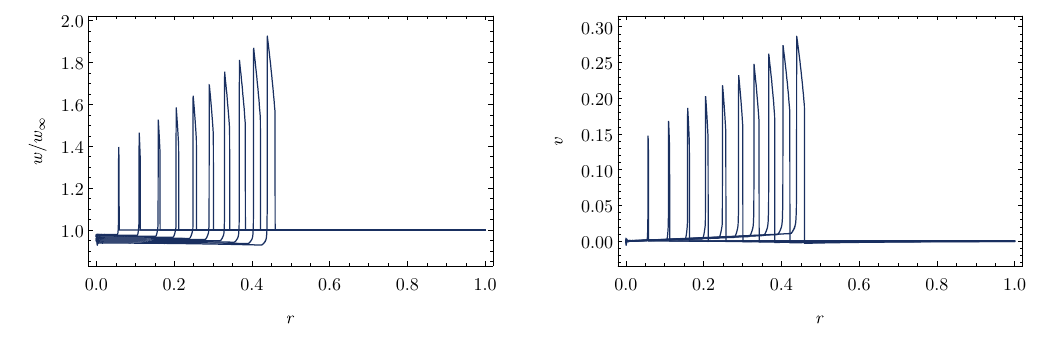}
\includegraphics[width=\columnwidth]{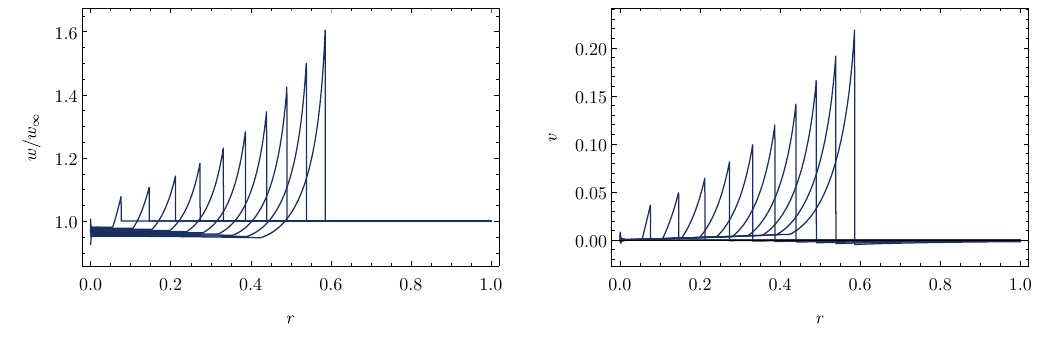}
\caption{\small
Fluid profile simulated at the parameter points similar to that in Fig.~\ref{fig:FRW}.
The radial coordinate is in the comoving unit and normalized by $[r]_{\rm sim} = H_0^{-1}$.
}
\label{fig:FRW2}
\end{center}
\end{figure}

Here we evolve the system for $n_t = 20000$ time steps with $[\Delta t]_{\rm sim} = 5 \times 10^{- 5} H_0^{- 1}$.
We take $n_r = 20000$ grid points and $[\Delta r]_{\rm sim} = 5 \times 10^{- 5} H_0^{- 1}$, with the same treatment of radial coordinate and boundary condition as before.
Note that $r$ is now in the co-moving unit and that the Hubble radius at initial locates $[r]_{\rm sim} \sim H_0^{- 1}$.
As in Fig.~\ref{fig:flat}, we set $v_w = (0.4, 0.6, 0.8)$ and $\alpha(t = 0) = \alpha_0 = 10^{-2}$, which grows to $\alpha(t = H_0^{- 1}) \sim 0.1$ at the end of simulation.
The results are shown in Fig.~\ref{fig:FRW2}. 
Note that we correctly reproduce the flat spacetime solutions for $t \ll H_0^{- 1}$.

These results have several characteristic features.
One is the breakdown of self-similarity of the profile due to the cosmic expansion.
As the latent heat fraction $\alpha(t)$ increases over time, the amplitude of the velocity and enthalpy also increases, forming sharper profiles than their flat-spacetime counterparts.
Another feature is the formation of the velocity tail behind the wall.
This feature can be understood as the motion of the fluid flowing into the empty region created by growing $\alpha(t)$:
as the value of $\alpha (t)$ increases, a larger proportion of the fluid around the wall gets swept outward, resulting in a relatively low density region just behind the wall.
Then the fluid in the inner region flows into this emptier region, forming the velocity tail of the fluid moving outward.
To check this interpretation, we ran simulations removing the $\Phi$ term from the equations by hand and confirmed that qualitatively similar tails still appear in the profile.
The last thing is the effect of the gravitational potential $\Phi$, which is barely visible in the current figure.
Careful look at the right bottom panel of Fig.~\ref{fig:FRW2} reveals slight inward velocity outside the bubble wall.
This effect, however, is rather negligible before the bubble grows to the horizon scale.
Therefore, we conclude that within this time scale, the most important gravitational effect on the fluid dynamics is the growing $\alpha(t)$ and the consequent formation of the velocity tail.

\begin{figure}[htbp]
\begin{center}
\includegraphics[width=0.48\columnwidth]{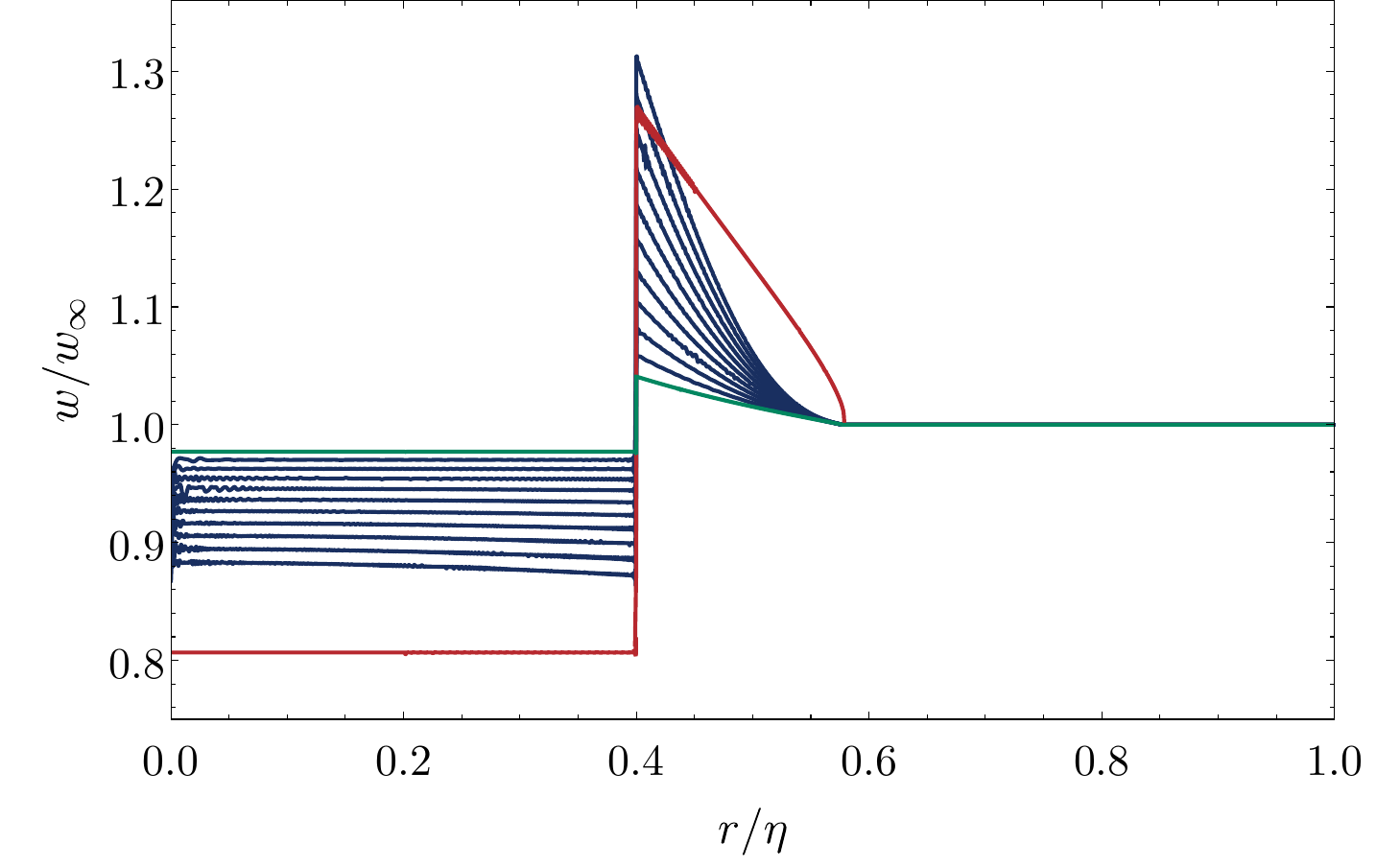}~
\includegraphics[width=0.48\columnwidth]{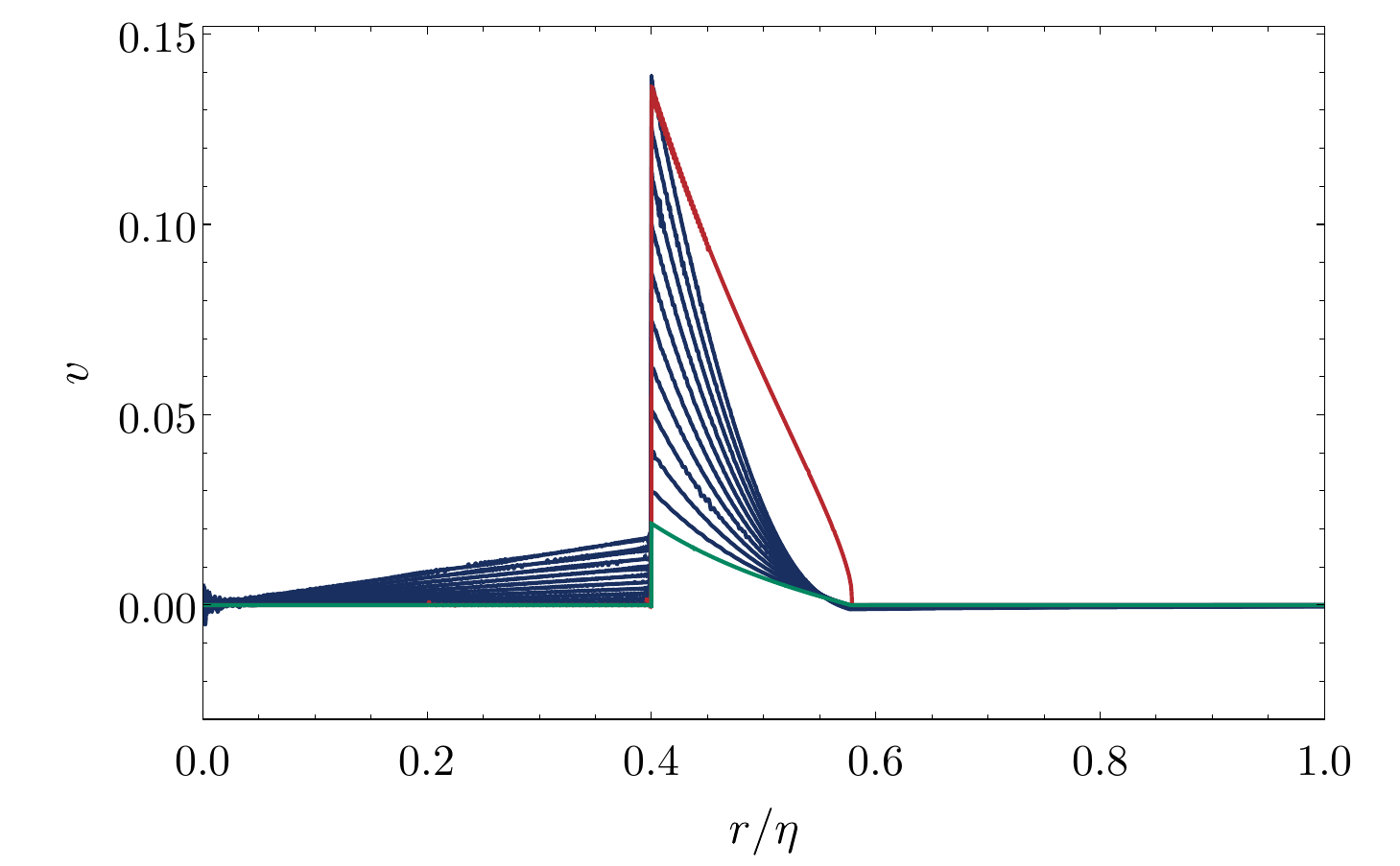}\\
\includegraphics[width=0.48\columnwidth]{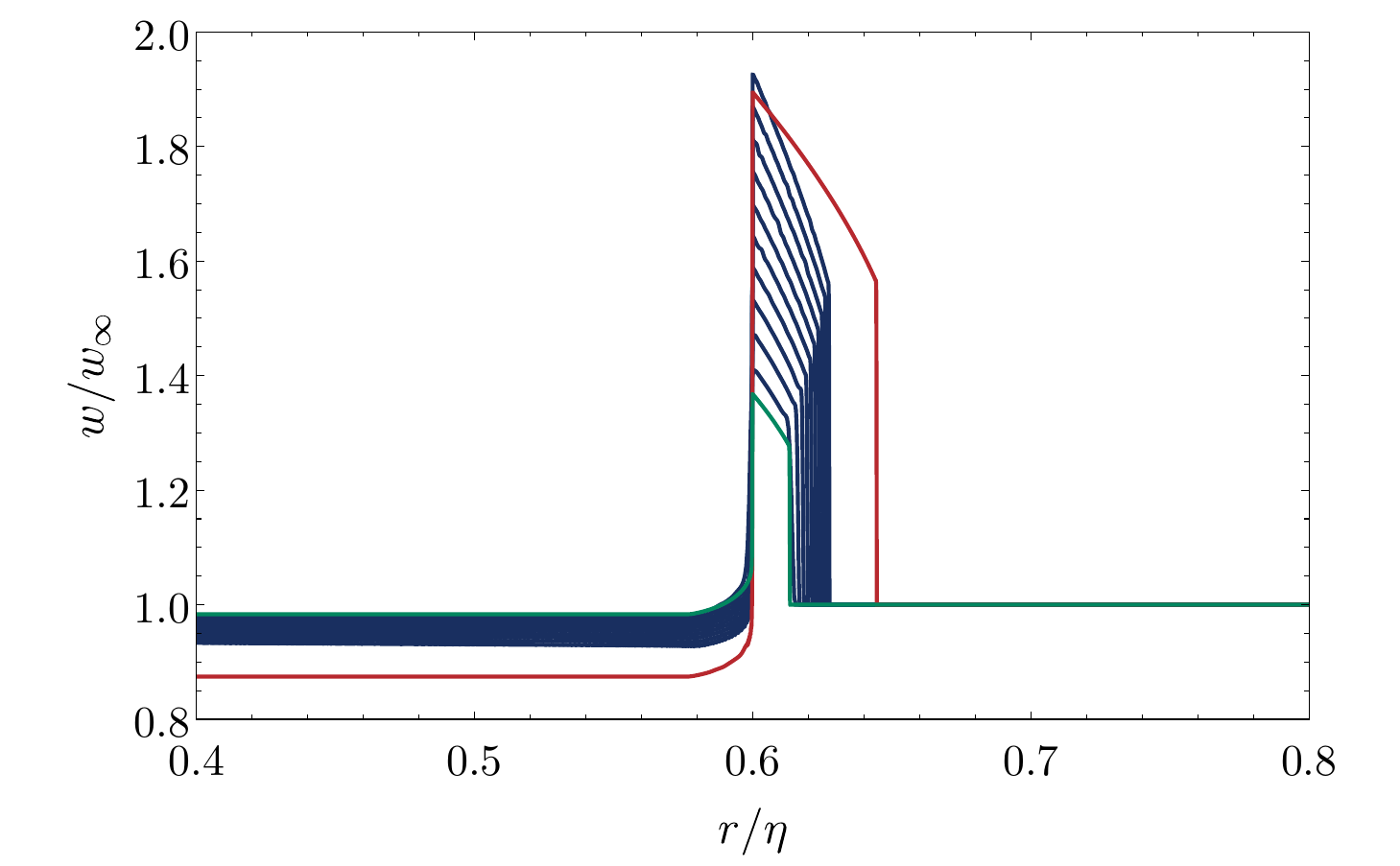}~
\includegraphics[width=0.48\columnwidth]{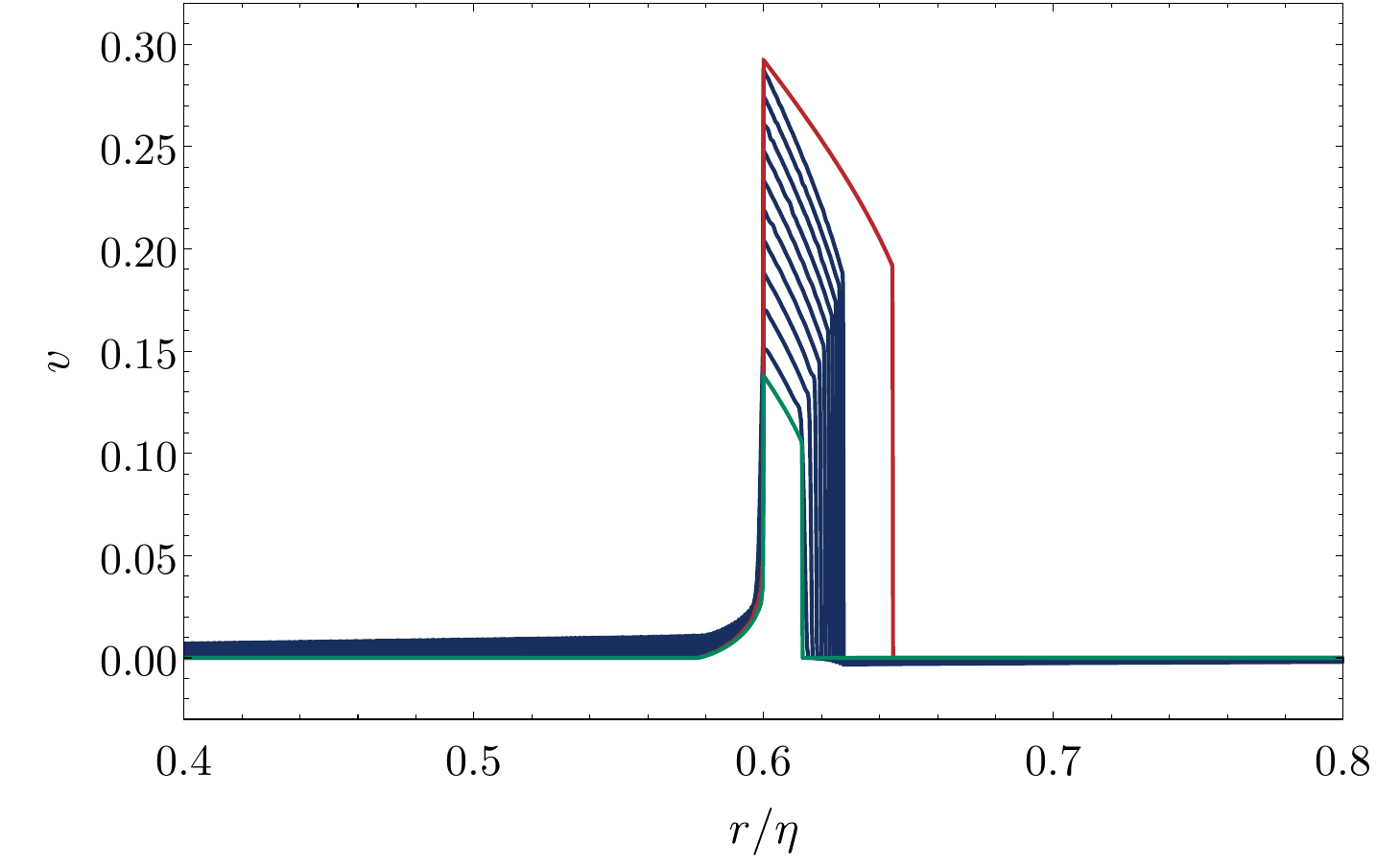}\\
\includegraphics[width=0.48\columnwidth]{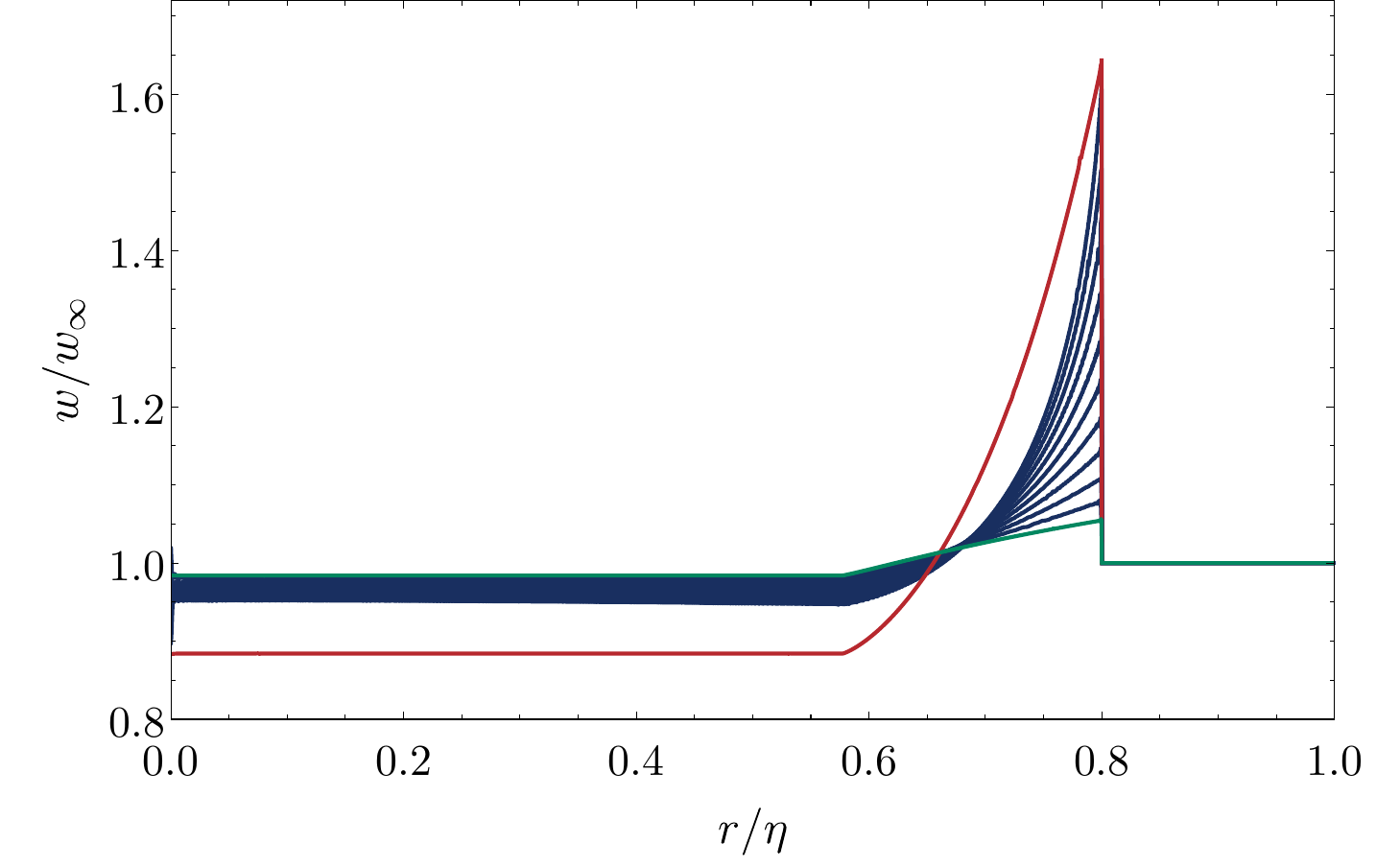}~
\includegraphics[width=0.48\columnwidth]{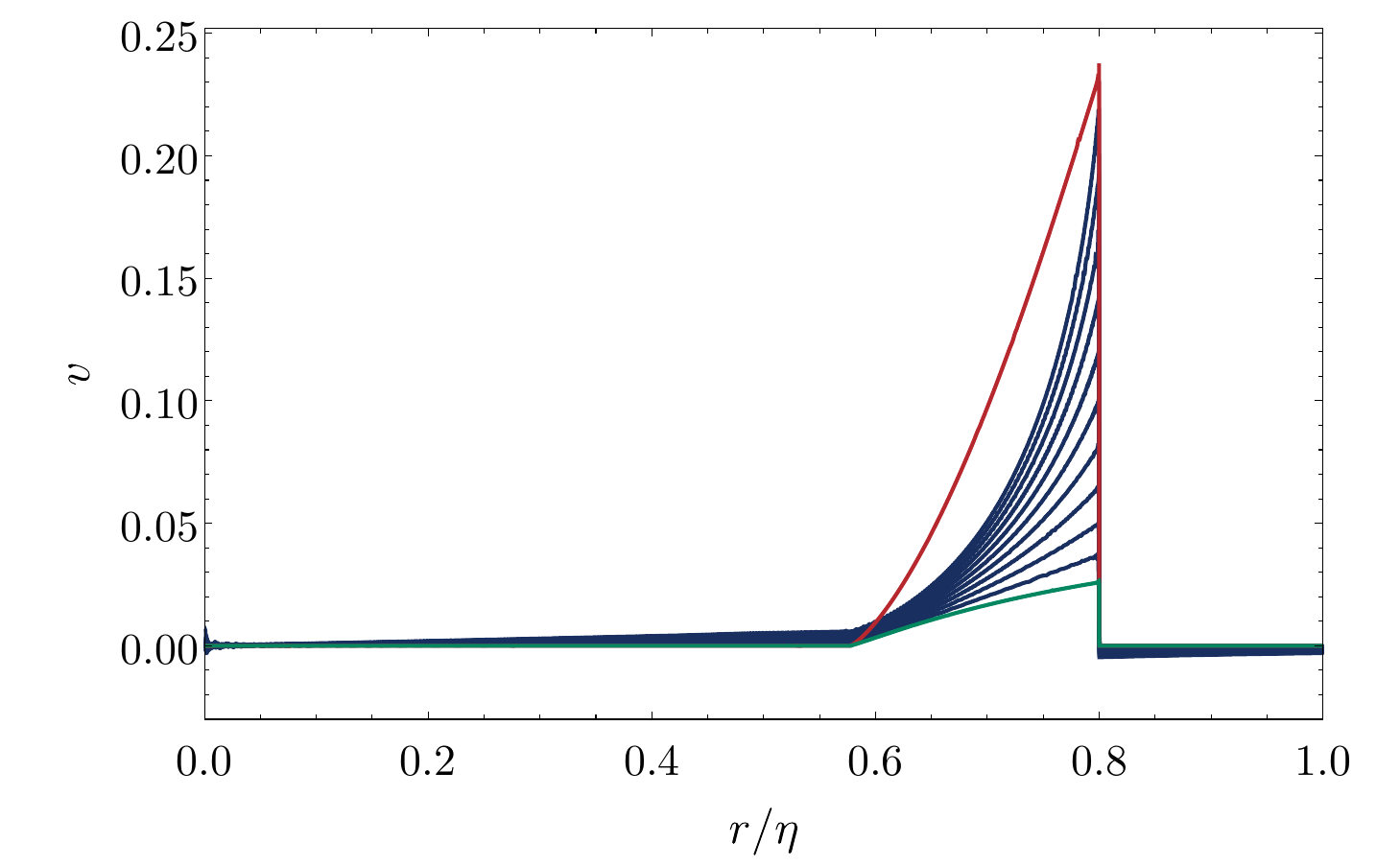}\\
\caption{\small
Enthalpy and velocity profile in self-similar coordinate for deflagration (top), hybrid (middle) and detonation (bottom).
Green and red lines correspond to the velocity profile in the flat spacetime background with $\alpha (H_0 t = 0)$ and $\alpha (H_0 t = 1)$, respectively.
}
\label{fig:desim}
\end{center}
\end{figure}

Now let us introduce a self-similar co-moving coordinate $r/\eta$, which was also used in Ref.~\cite{Cai:2018teh}. As expected from Eq.~\eqref{eq:wall_position}, this normalization is useful for comparing our result with the flat spacetime solutions. 
In Fig.~\ref{fig:desim}, ``non self-similar evolution'' of the fluid velocity is plotted in the self-similar coordinate for the set of parameters $(\alpha_0,  v_w) = (10^{-2}, 0.4), (10^{-2}, 0.6), (10^{-2}, 0.8)$, from the top panel to the bottom one. 
For each plot, self-similar solutions in the flat spacetime with $\alpha = \alpha(t = 0)$ (green) and $\alpha = \alpha(t = H_0^{-1})$ (red) are plotted.
In each case, one can see that over the time scale $t \sim H_0^{-1}$, the fluid eventually develops a thinner profile than the corresponding solution in the flat spacetime. 
This feature, which would affect the spectrum of SGWB from sound wave, was first pointed out in Ref.~\cite{Cai:2018teh} studying the late-time ($t \gg H_0^{-1}$) hydrodynamic solution in FLRW background. 
Our result indicates that this thinning is also true for the intermediate time scale where the bubble size is still of sub-Hubble scale.

\subsubsection{Late time solutions and the efficiency factor}\label{sec:late}
In the early-time solutions we see the importance of the cosmological expansion.
Here we perform simulations for a longer period to see the asymptotic behavior of the system until $\alpha (t)$ becomes of order unity and $\Phi$ becomes sizable.
We solve the system until max$(v(t, r))$ reaches $v_w$ or $\Delta\Phi(t) \equiv |\Phi(t,0) - \Phi(t, r\to\infty)|$ reaches $0.5$. 
Too large $\alpha(t)$ such that max$(v(t, r)) > v_w$ indicates the breakdown of the assumption about constant wall velocity, and we investigate accelerating walls in the next subsection.

\begin{figure}[htbp]
\begin{center}
\includegraphics[width=\columnwidth]{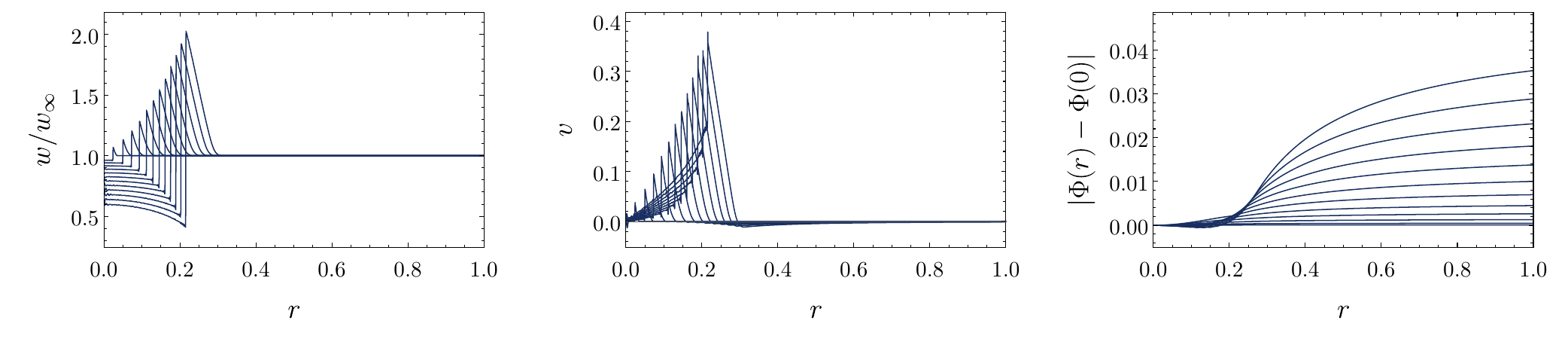}\\
\includegraphics[width=\columnwidth]{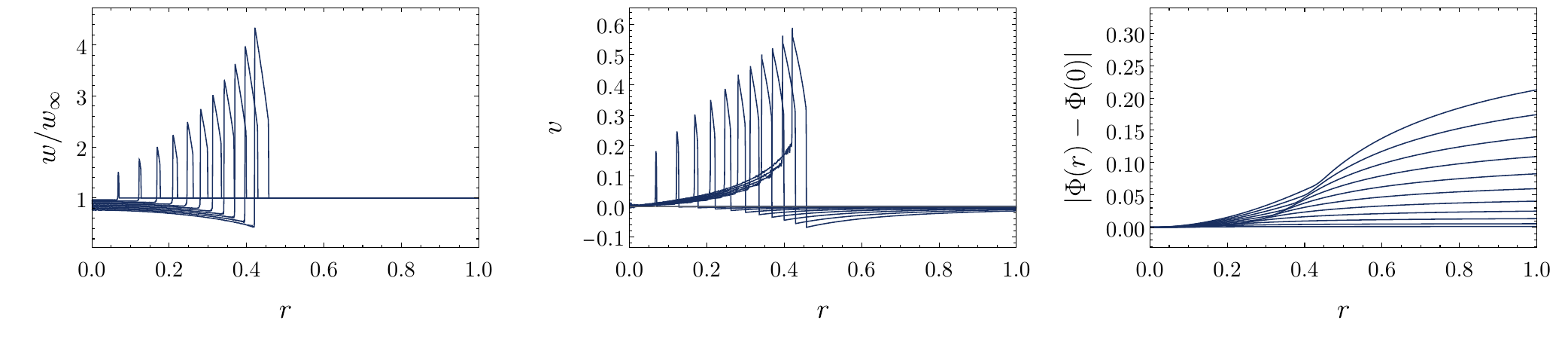}\\
\includegraphics[width=\columnwidth]{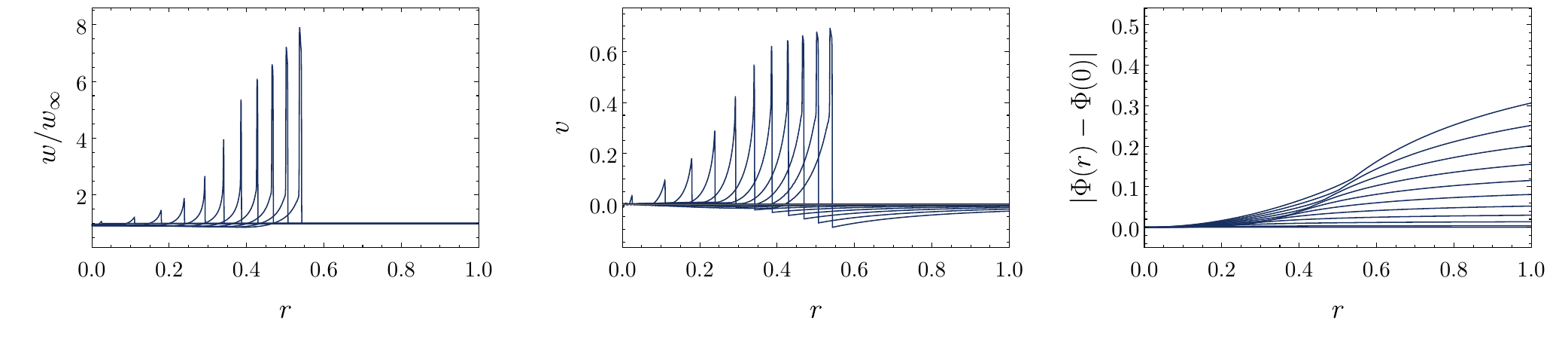}\\
\includegraphics[width=\columnwidth]{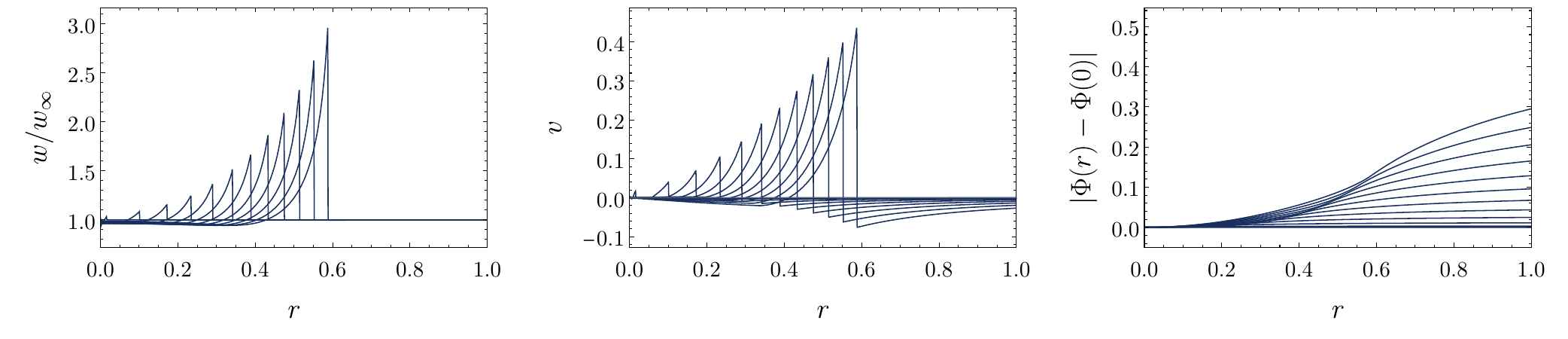}
\caption{\small
Late time profile of enthalpy, velocity and gravitational potential (from left to right column). Here again we set $\alpha_0 = 10^{-2}$ and $v_w = (0.4,0.6,0.8,1)$ (from top to bottom raw). Here the radial coordinate is normalized by $\left[ r \right]_{\rm sim} = 3H_0^{-1}$, which is 3 times larger than that in Fig.~\ref{fig:FRW2}}
\label{fig:FRW_late}
\end{center}
\end{figure}

The results of the simulation are shown in Fig~\ref{fig:FRW_late}.
In this figure, we solve the system for the same parameter values as in Fig~\ref{fig:FRW2}.
To evaluate $\Delta\Phi(t)$, we increase the box size by lowering the spatial resolution to $[\Delta r]_{\rm sim} = 10^{-3} H_0^{-1}$ while keeping $n_r = 20000$.
We also set $[\Delta t]_{\rm sim} = 5 \times 10^{-4} H_0^{-1}$ to facilitate a longer time scale in the simulation.
Note that the late-time behavior of the system is not much affected by the reduced resolution.
This is because the late-time behavior is dominantly determined by the late-time energy injection: 
though the spatial resolution at initial times is lower than what is needed to fully capture the shell profile, the error at these times is negligible. The asymptotic profile is determined by the late-time energy injection by large $\alpha (t)$, when the bubble and fluid shell has grown large enough to be captured by the current spatial resolution.
Here we also add the luminal case $v_w = 1$ since the wall velocity typically reaches to the speed of light in the scenarios of extremely supercooled FOPT.

We observe an interesting time evolution in the third example of Fig~\ref{fig:FRW_late} where $v_w = 0.8$. 
The fluid profile is in the detonation regime according to $(\alpha_0, v_w) = (10^{-2}, 0.8)$ at earlier times, while at later times the larger value of $\alpha(t)$ drives the fluid profile into the hybrid regime.
This behavior is indeed expected from the phase diagram of flat spacetime solutions depicted in, for example, Fig.~7 of Ref.~\cite{Espinosa:2010hh}. When $v_w > c_s$, as $\alpha$ is increased from a small value to a large value, one can see that the profile transitions from the detonation region to the hybrid region. As $\alpha(t)$ increases with time in our case, such a transition is expected to inevitably occur depending on the initial conditions. Therefore, we conclude that such a transition from detonation to hybrid solution is a general behavior for the cosmological time scale evolution.

In this figure we also plot the spatial profile of gravitational potential $\Phi(t,r)$ as it starts to give a non-negligible contribution.
At large $r$, we found that $\Phi(t,r)$ decays $1/r$ as expected.
This time one clearly sees that $\Phi(t,r)$ causes the inward velocity outside the wall without disturbing the enthalpy there.\footnote{Only for $v_w = 0.4$, the simulation was stopped due to max$\left( v(t,r) \right) = v_w$ and the gravitational potential was yet to grow as the same order of the other examples.}
The dominant contribution to this inward velocity comes from the second term in Eq.~(\ref{eq:eom_grav_pot}) that represents the absence of the vacuum energy inside the bubble.
We confirmed that this inward velocity is still present, albeit reduced, even if we solve the system with the second term in Eq.(\ref{eq:eom_grav_pot}) set to $0$ by hand.
This fact indicates that the highly concentrated fluid shell starts to work as the source of gravitational attraction for the fluid outside the wall.
In any case, in estimating the energy budget below, we find that the contribution from the fluid inward motion is still energetically sub-dominant, as the fluid enthalpy is sharply peaked around the bubble wall.

As we have seen the behavior of the late time solutions with specific parameters, now let us quantitatively investigate the solutions for various sets of input parameters $v_w$ and $\alpha_0$.
For this purpose, here we evaluate the kinetic energy fraction defined as
\begin{equation}
    \kappa(t) = \frac{4}{r_w^3(t) \alpha(t)w_b(t)}\int dr r^2 wv^2\gamma^2.
    \label{eq:kinetic}
\end{equation}
By definition, this quantifies the fraction of released energy transferred into the fluid kinetic energy, allowing one to estimate the strength of SGWB sourced by the sound waves.
Since our solutions develop a thinner fluid shell than that of the self-similar solution in the flat spacetime, the efficiency factor at late time $t$ should be smaller than that of self-similar solution given by $v_w$ and $\alpha(t)$.
Quantifying this suppression helps us to give a better estimate of the sound wave contribution for the slower FOPT.

\begin{figure}[htbp]
\begin{center}
\includegraphics[width=0.7\columnwidth]{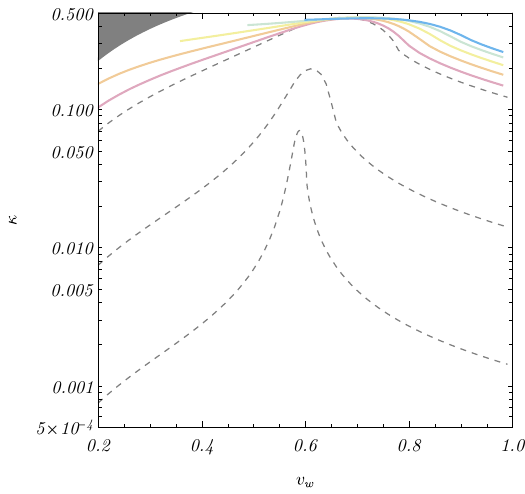}
\vskip 0.5cm
\includegraphics[width=0.7\columnwidth]{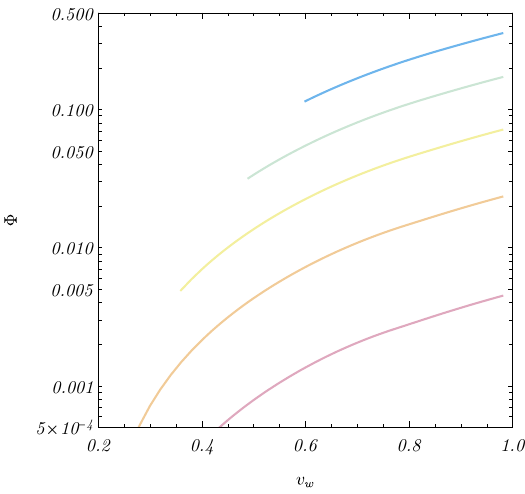}
\caption{\small
Contours of $\alpha(t)$ in $(v_w, \kappa)$-plane (top) and in $(v_w, \Phi)$-plane (bottom).
The initial value of $\alpha$ is set as $\alpha_0 = 10^{-1}$. The contour values are logarithmically spaced by $10^{-0.2}$, starting from $10^{-0.8}$.
Dashed lines in the top panel is $\kappa$ computed with flat space time solutions for $\alpha = 10^{-3}, 10^{-2}, 10^{-1}$.
The shaded are represents the kinetically prohibited region for the flat spacetime solution.}
\label{fig:contour1}
\end{center}
\end{figure}

\begin{figure}[htbp]
\begin{center}
\includegraphics[width=0.7\columnwidth]{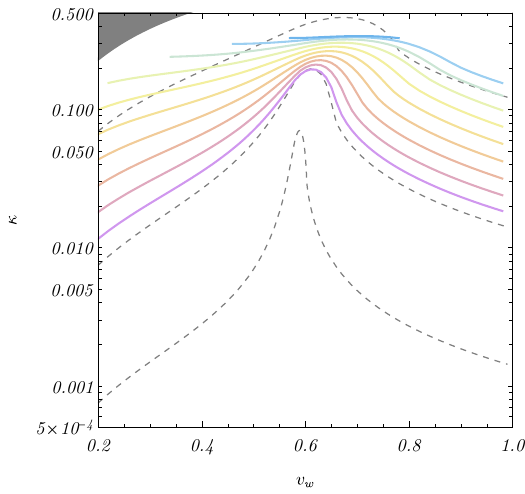}
\vskip 0.5cm
\includegraphics[width=0.7\columnwidth]{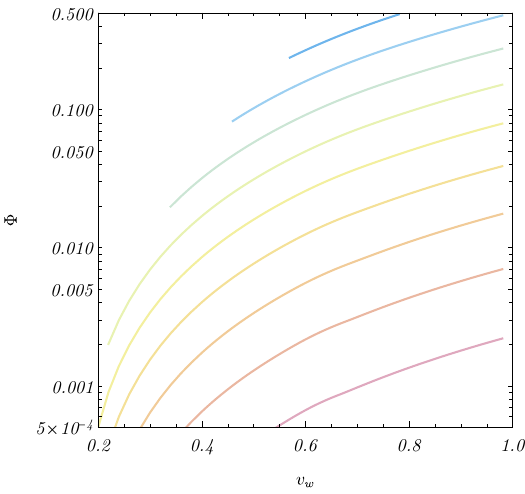}
\caption{\small
The same contour plots as Fig.~\ref{fig:contour1} but for $\alpha_0 = 10^{-2}$. The contour values are logarithmically spaced by $10^{-0.2}$, starting from $10^{-1.8}$.
}
\label{fig:contour2}
\end{center}
\end{figure}

\begin{figure}[htbp]
\begin{center}
\includegraphics[width=0.7\columnwidth]{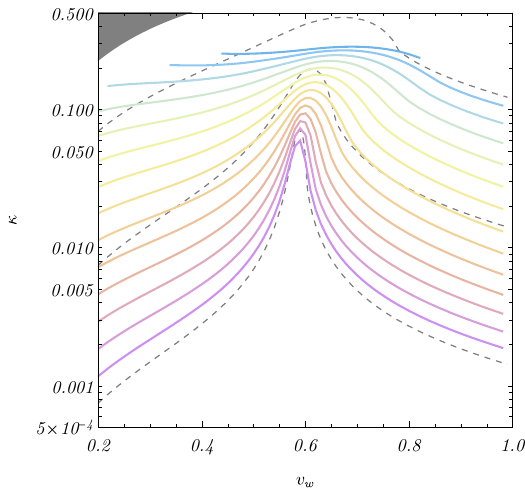}
\vskip 0.5cm
\includegraphics[width=0.7\columnwidth]{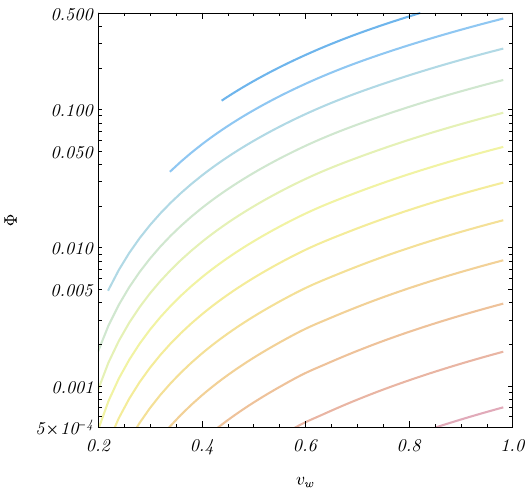}
\caption{\small
The same contour plots as Fig.~\ref{fig:contour1} but for $\alpha_0 = 10^{-3}$. The contour values are logarithmically spaced by $10^{-0.2}$, starting from $10^{-2.8}$.
}
\label{fig:contour3}
\end{center}
\end{figure}

For $\alpha_0 = 10^{-1}, 10^{-2}$, $10^{-3}$ and different values of $v_w \in [0.2, 1.0]$, we evolved the system and evaluate $\kappa(t)$ until max$(v(t, r))$ reaches $v_w$ or $\Delta\Phi(t)$ reaches $0.5$.
For the evaluation of $\Delta\Phi(t)$, we use the value of $\Phi(t,r)$ at $r = [r_{\rm max}]_{\rm sim}$ and use the following approximation
\begin{equation}
    \Phi(t,r) \sim \Phi(t,r_w)\frac{r_w}{r},
\end{equation}
which well holds for $r > r_w$ since the main contributions to $\Phi$ are well localize at $r < r_w$.
To have a better accuracy in this evaluation, we set the number of spatial grids for $v_w \geq 0.6$ so that $[r_{\rm max}]_{\rm sim} = 3r_w(t)\vert_{\alpha(t) = 1}$. 
For $v_w < 0.6$, we use the value of $[r_{\rm max}]_{\rm sim}$ at $v_w = 0.6$ since the leading edge of fluid shell moves at $v = c_s \sim 0.6$ even for $v_w < c_s$.
Here we also set higher resolution $[\Delta r]_{\rm sim} = [\Delta t]_{\rm sim} = 5\times10^{-5}H_0^{-1}$ to capture the thinnest shock structure around the beginning of the simulation as much as possible.
The results are shown in the top panels of Figs.~\ref{fig:contour1}--~\ref{fig:contour3} where the contours of $\alpha(t)$ are drawn on $(v_w, \kappa(t))$ plane.\footnote{We should note that in the plots for $\alpha_0 = 10^{-3}, 10^{-2}$, the first one or two contours (or time slices) were yet to completely capture the shock profile, especially for the wall velocity around $v_w \sim 0.6$. 
But it is suffice for our purpose to compare the behavior at later time.}
We also draw $\alpha(t)$ contours on $(v_w, \Delta\Phi(t))$ plane as shown in the bottom panels of Figs.~\ref{fig:contour1}--~\ref{fig:contour3}.
Notice that the monotonic increase of $\Delta\Phi(t)$ against $v_w$ is reasonable as the bubble becomes larger as $v_w$ increases. 

The top panels can be compared to Fig.~8 in Ref.~\cite{Espinosa:2010hh}, which evaluates the kinetic energy fraction for the flat spacetime solutions.
For the reference, we plot $\kappa$ for those with $\alpha = 10^{-3}, 10^{-2}, 10^{-1}$.
We find that in our case, the growth of $\kappa(t)$ with respect to the increase of $\alpha(t)$ is quite inefficient compared to the self-similar solution.
Recalling that the effect of gravitational potential seems to be energetically negligible, the thinner profile of fluid shell mostly accounts for this suppression.

\begin{figure}[htbp]
\begin{center}
\includegraphics[width=0.45\columnwidth]{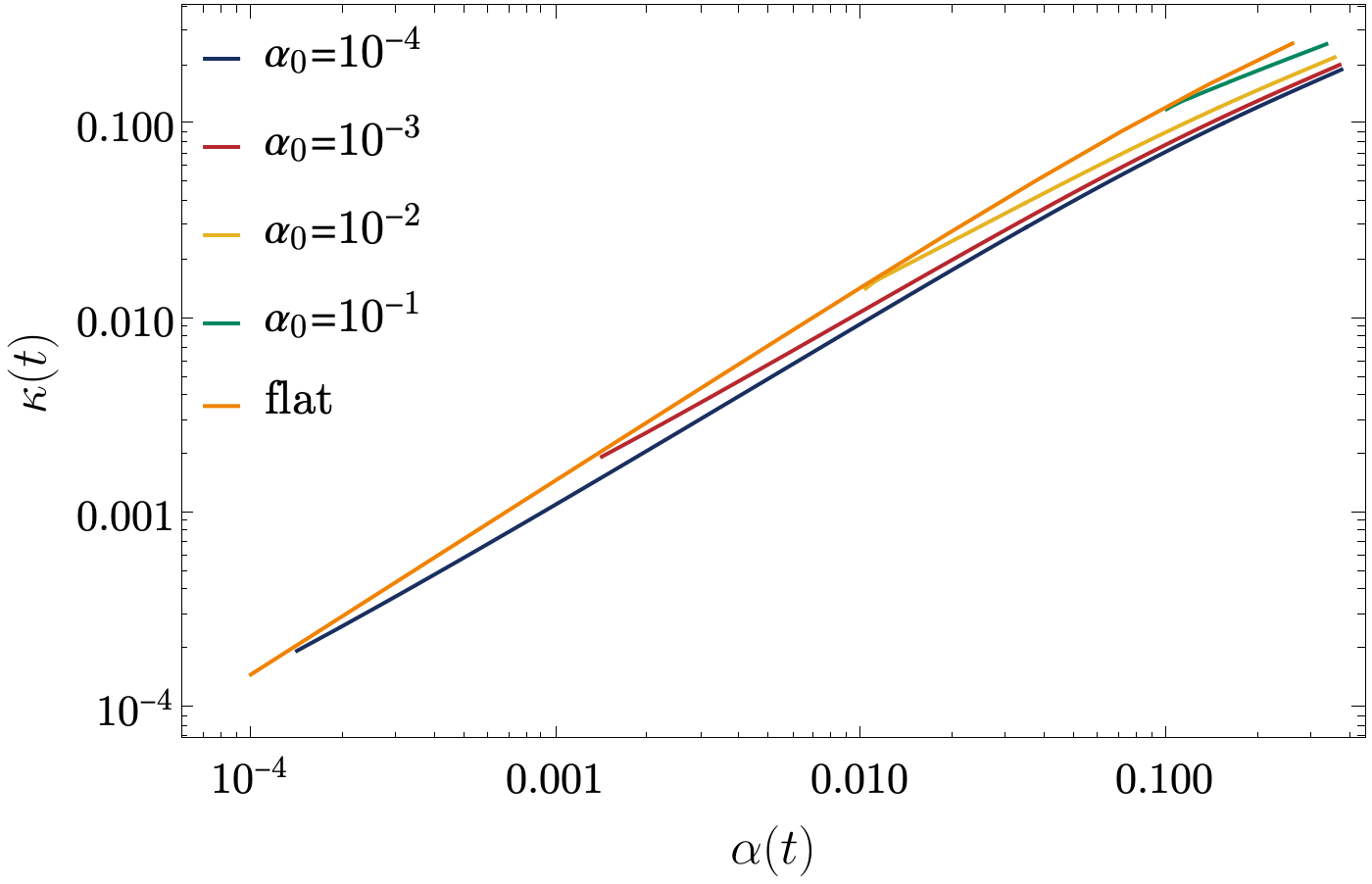}
\includegraphics[width=0.45\columnwidth]{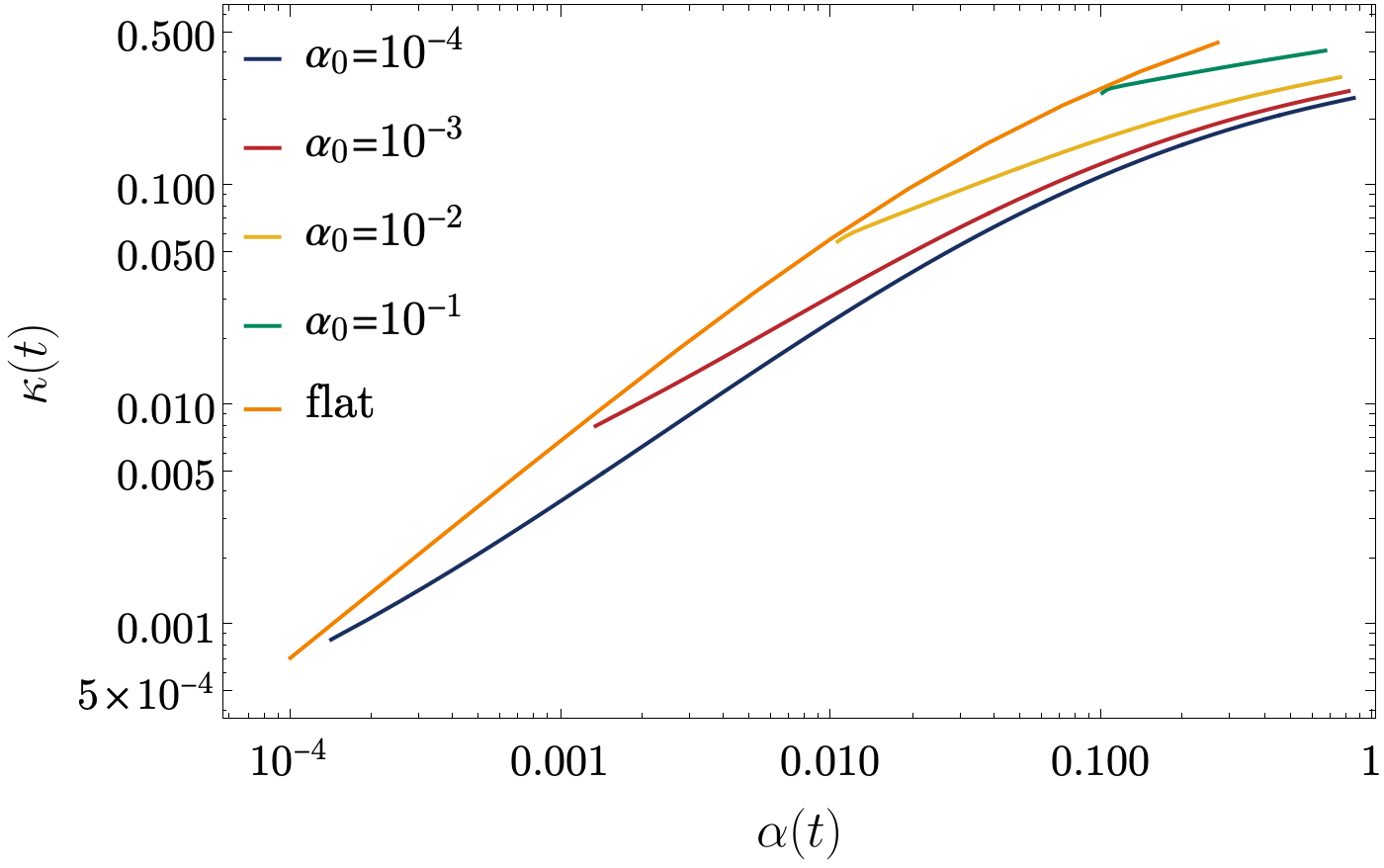}
\\[0.5cm]
\includegraphics[width=0.45\columnwidth]{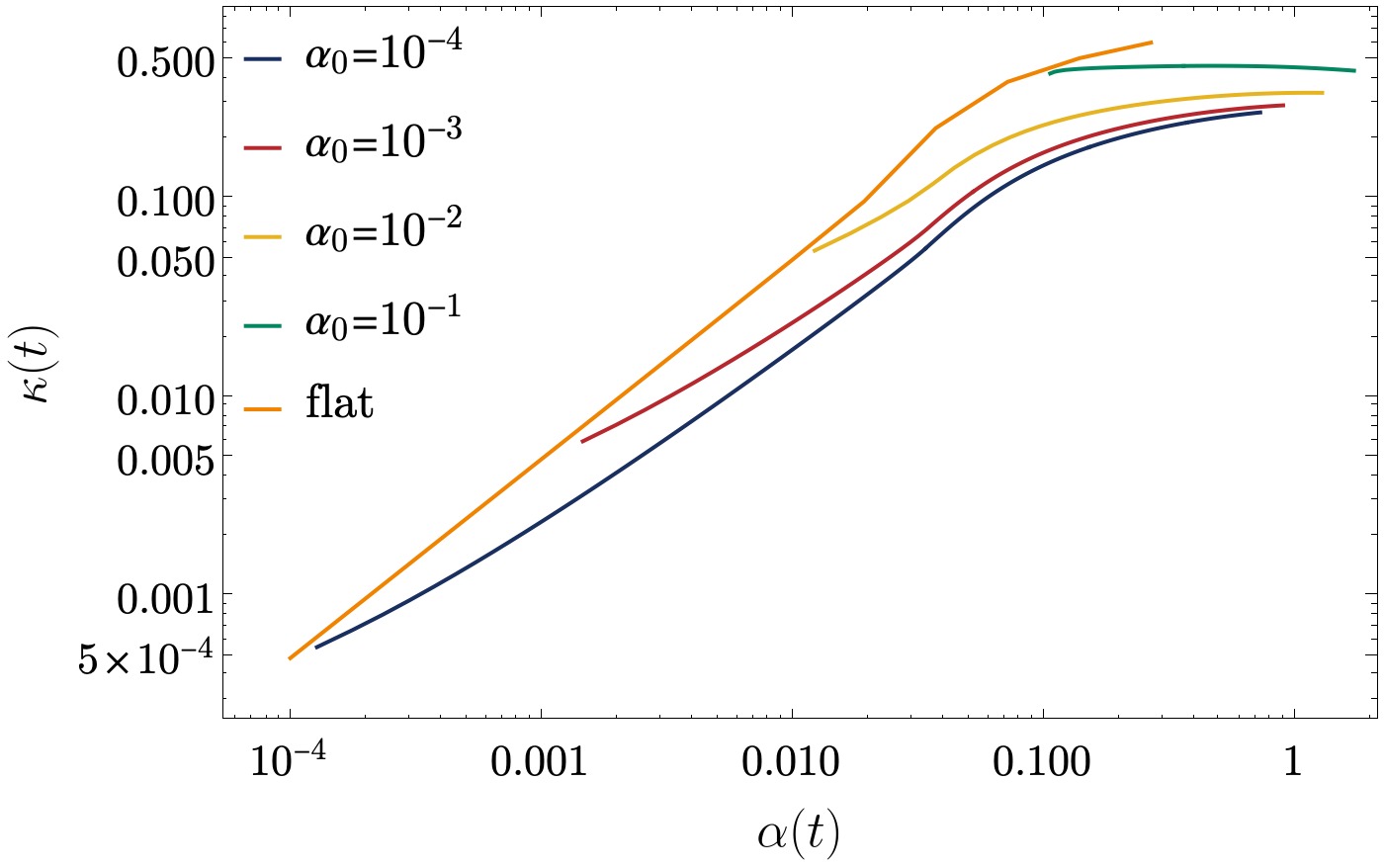}
\includegraphics[width=0.45\columnwidth]{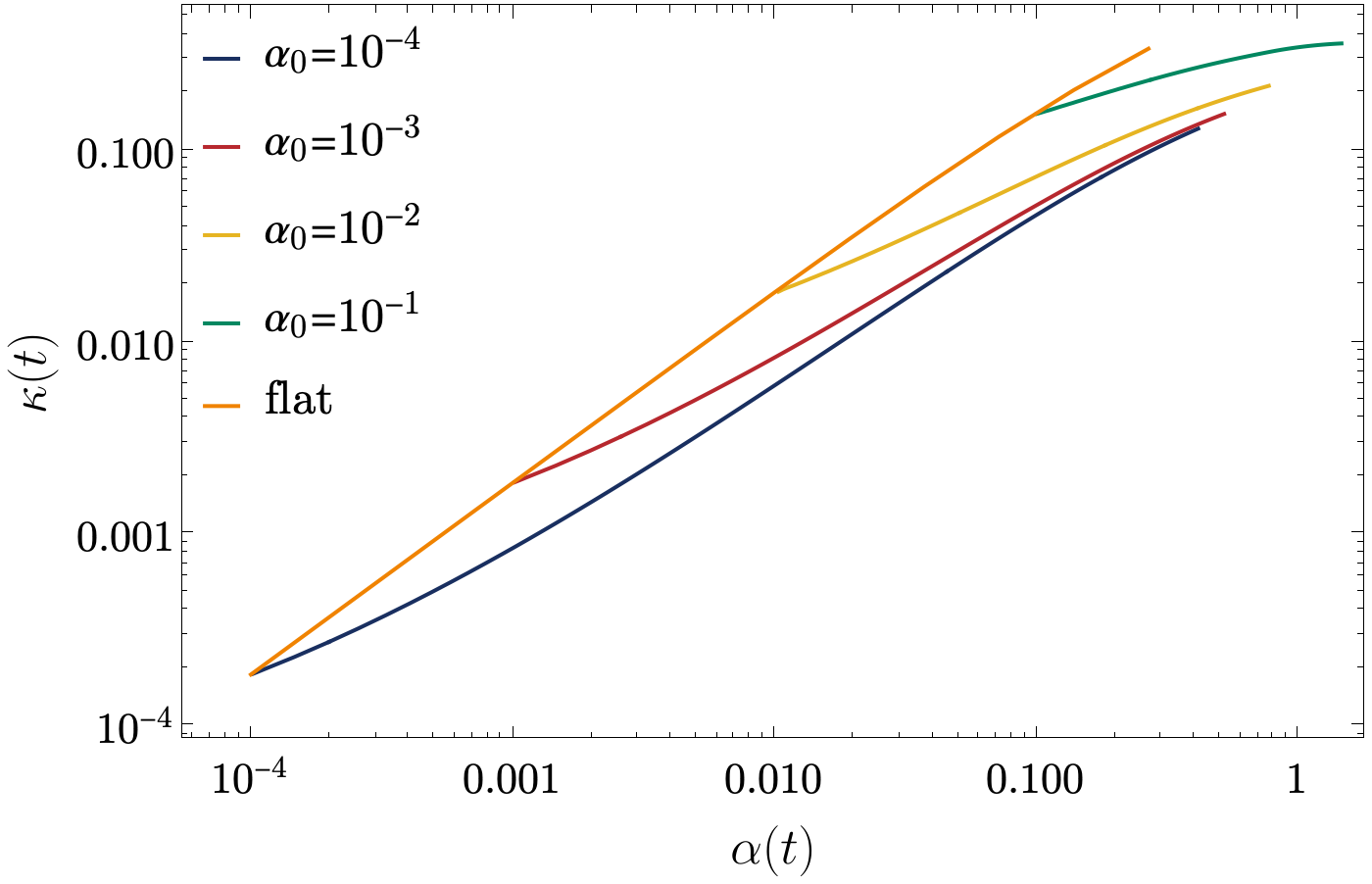}
\caption{\small
The evolution of $\kappa(t)$ against the different initial condition $\alpha_0$ are shown for the wall velocity $v_w = 0.3, 0.5, 0.7, 0.9$ (top-left, top-right, bottom-left, bottom-right respectively). 
For comparison, $\kappa$ of self-similar solution in the flat spacetime is plotted for values of $\alpha$.
}
\label{fig:kappa_conv}
\end{center}
\end{figure}

We also found that the late time magnitude of $\kappa(t)$ (and also $\Delta\Phi(t)$) seems to converge when $\alpha_0$ decreases, by comparing Figs.~\ref{fig:contour2} and~\ref{fig:contour3}. 
To investigate this behavior, we also simulated the system for $\alpha_0 = 10^{-4}$ with specific values of $v_w = 0.3, 0.5, 0.7, 0.9$.
In Fig.~\ref{fig:kappa_conv}, the evolution of $\kappa(t)$ is shown for different wall velocities and $\alpha_0 = 10^{-1}, 10^{-2}, 10^{-3}, 10^{-4}$.
One can understand that the larger value of $\kappa(t)$ for larger $\alpha_0$ is because the shell is still in the process of becoming thinner.
In either wall velocity, the degree of decrease in the value of $\kappa(t)$ becomes smaller as $\alpha_0$ decreases.
This could be consistent with Ref.~\cite{Cai:2018teh} that derives the late-time limit solution, indicating the existence of a limit to the thinning of fluid shell.
We believe that this convergent behavior is reasonable since for $\alpha_0$ small enough, late time dynamics driven by the horizon-scale bubble should become independent of precisely when the bubble nucleates.
Here again we plot the correspondent value of $\kappa$ for the flat spacetime solutions and one can read to what extent $\kappa(t)$ is suppressed over the cosmological evolution.\footnote{For $v_w = 0.5$ and $v_w = 0.7$, the shock profile is not completely captured in the earliest time steps and this results in a slight deviation from the flat spacetime case. Notice that this is not a physical deviation, but the deviation in the later time steps is certainly caused by the cosmological expansion.}
For interested readers, we also present the similar plot for different physical quantities in App.~\ref{app:late_time}.

\subsection{Accelerating walls}\label{sec:wall_acc}

Finally, we consider a situation where the wall velocity admits a time dependence.
Such a dependence arises because the cosmic expansion dilutes the radiation coupled to the wall (or the scalar field). Consequently, the fluid pressure should decrease over time and the wall velocity rapidly approaches $v_w = 1$ (see, for example, Fig.~10 of Ref.~\cite{Espinosa:2010hh}). 
Microphysically, the wall motion is governed by the equation of motion
\begin{align}
\Box \phi + \frac{\partial V_0}{\partial \phi} + \sum_i \frac{d m_i^2}{d \phi} \int \frac{d^3 p}{(2 \pi)^3 2 E}~f_i (p)
&=
0.
\end{align}
The distribution function can be decomposed into the equilibrium part and deviation from it as
\begin{align}
f_i (p)
&=
f^{\rm eq}_i (p) + \delta f_i (p),
\end{align}
and the former is combined with the zero-temperature potential to give the free energy.
The latter gives the friction term
\begin{align}
{\cal K} (\phi)
&=
\sum_i \frac{d m_i^2}{d \phi} \int \frac{d^3 p}{(2 \pi)^3 2 E}~\delta f_i (p).
\end{align}
Phenomenological modelings of the friction adopted in Refs.~\cite{Ignatius:1993qn,Megevand:2009gh} and in Ref.~\cite{Espinosa:2010hh} are respectively
\begin{align}
{\cal K} (\phi)
&\propto
\left\{
\begin{array}{cc}
\displaystyle
T u^\mu \partial_\mu \phi
&\qquad
{\rm model~1},
\\[0.2cm]
\displaystyle
T \frac{u^\mu \partial_\mu \phi}{\sqrt{1 + (\lambda_\mu u^\mu)^2}}
&\qquad
{\rm model~2},
\end{array}
\right.
\end{align}
with $\phi$ being the scalar field driving the transition and $\lambda_\mu = (0, 0, 0, 1)$ in the wall frame (with the wall motion taken to be in the $z$ direction).
Temperature $T$ here corresponds to the one deep inside the symmetric phase.
The latter modeling was originally motivated by the fact that the friction reaches a maximum in the $v_w \to 1$ limit~\cite{Bodeker:2009qy}, while it is now known that particle splitting processes lead to friction terms increasing with $\gamma_w = 1 / \sqrt{1 - v_w^2}$ deep inside the relativistic regime~\cite{Bodeker:2017cim} (see also Refs.~\cite{
Hoche:2020ysm,Azatov:2020ufh,Gouttenoire:2021kjv,Azatov:2023xem,Long:2024sqg}).
However, the friction from particle splitting becomes important only after the wall gets almost luminal, so it is expected that at the initial stage of bubble expansion the friction grows linearly in $v_w$, and thus in this paper we consider both cases.
For model 1 and 2, we set $\gamma_w v_w \propto \alpha$ and $v_w \propto \alpha$ respectively, and get the expressions
\begin{align}
v_w(t)
&=
\left\{
\begin{array}{cc}
\displaystyle
\sqrt{\frac{\alpha^2 \gamma_w^2 (t = 0) v_w^2 (t = 0)}{\alpha^2_0 + \alpha^2 \gamma_w^2 (t = 0) v_w^2 (t = 0)}}
&\qquad
{\rm model~1},
\\[0.5cm]
\displaystyle {\rm min} \left[ \frac{\alpha(t)}{\alpha_0} v_w (t = 0), 1 \right]
&\qquad
{\rm model~2}.
\end{array}
\right.
\label{eq:toy_model}
\end{align}

\begin{figure}[htbp]
\begin{center}
\includegraphics[width=0.5\columnwidth]{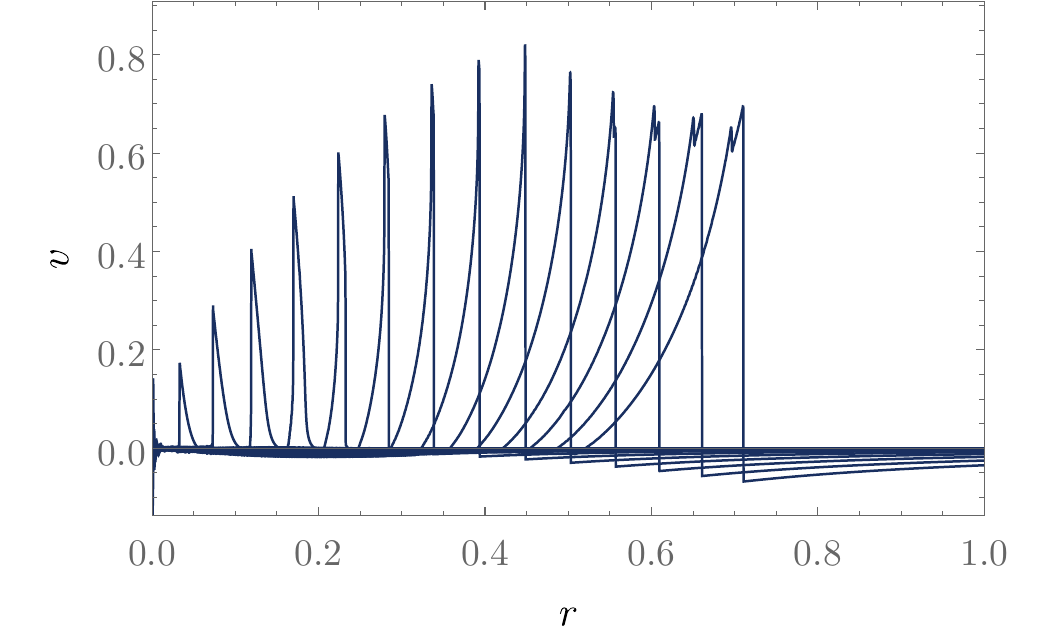}~
\includegraphics[width=0.5\columnwidth]{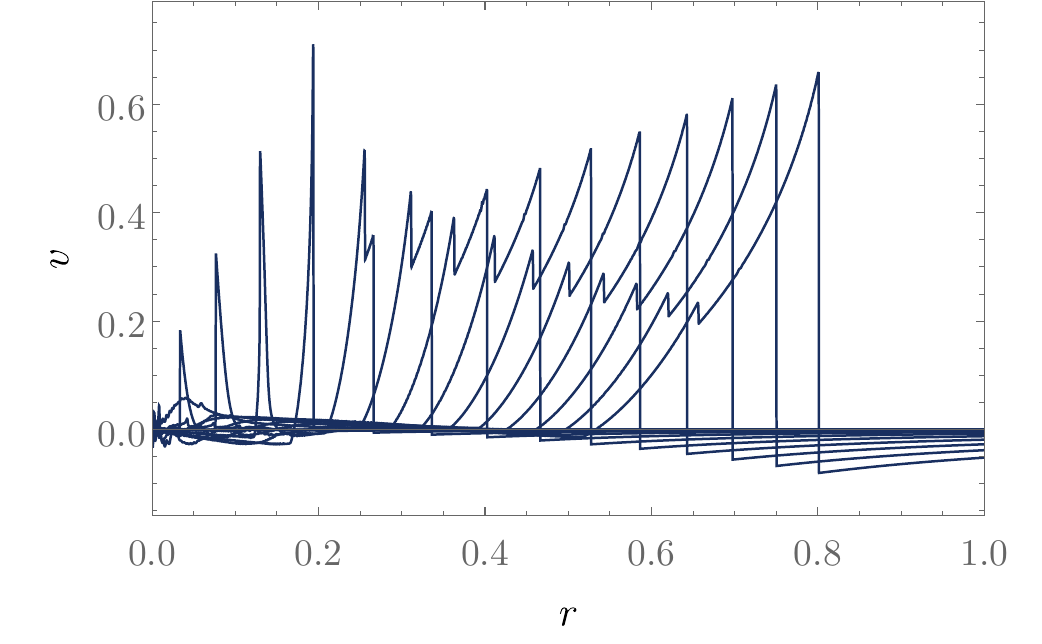}\\
\includegraphics[width=0.5\columnwidth]{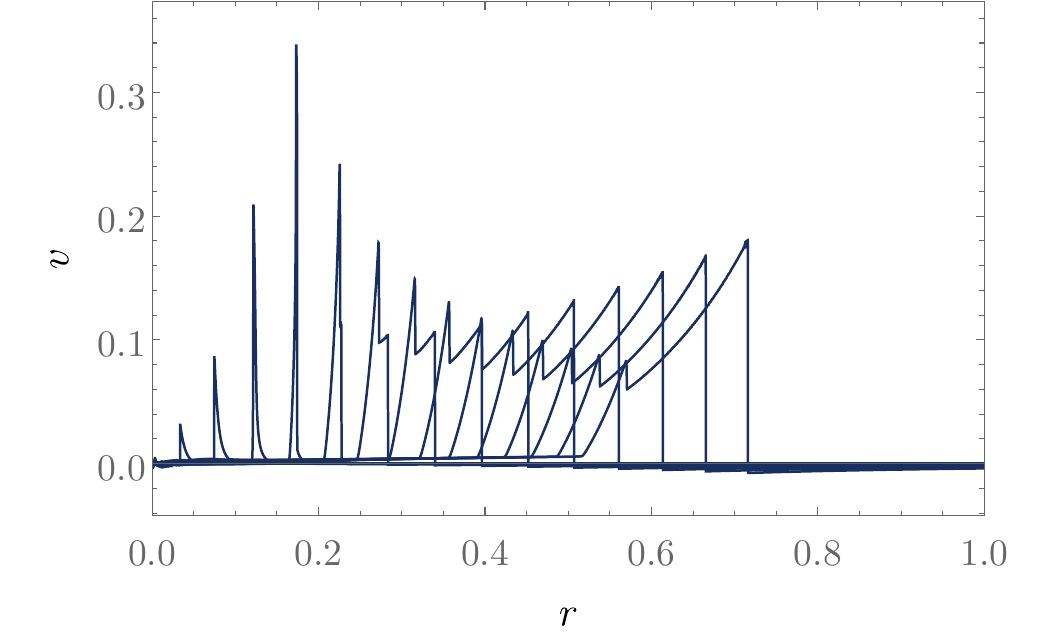}~
\includegraphics[width=0.5\columnwidth]{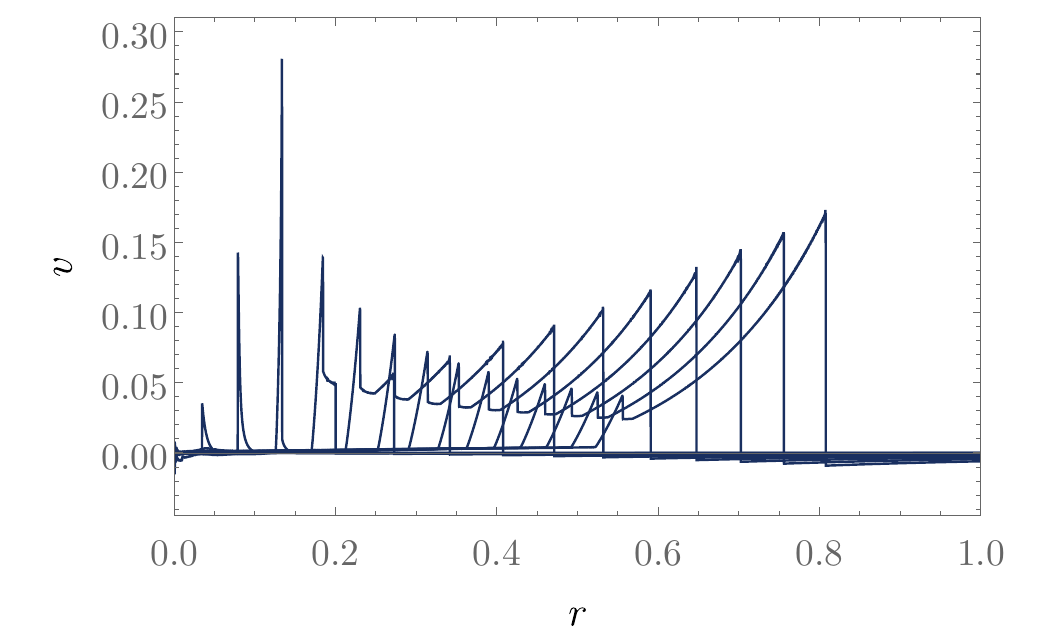}
\caption{\small
Velocity profiles simulated with allowing the wall velocity to increase over time as Eq.~\eqref{eq:toy_model}. 
The left panels are for model~1 and the right panels are for model~2. Top raw is evolved from $(v_w(t = 0), \alpha_0) = (3 \times 10^{-1}, 10^{-1})$ and the bottom raw is from $(v_w(t = 0), \alpha_0) = (3 \times 10^{-1}, 10^{-2})$ while $a(t) \leq 2$.
Here the radial coordinate is normalized by $\left[ r \right]_{\rm sim} = 2H_0^{-1}$.
Except for the left top panel, one can see the formation of substructure, which originates from the discontinuity in the maximum velocity between the hybrid and detonation regime.}
\label{fig:moving_wall}
\end{center}
\end{figure}

In the left and right panels of Fig.~\ref{fig:moving_wall}, we show the velocity profile of fluid for model~1 and model~2, respectively.
Here we take the same resolution as that in Sec.~\ref{sec:comparison_flat}.
The system is evolved until $a(t) \leq 2$ in all cases and the initial condition was set as $(v_w(t = 0), \alpha_0) = (3 \times 10^{-1}, 10^{-1}) $ for the top raw and $(v_w(t = 0), \alpha_0) = (3 \times 10^{-1}, 10^{-2})$ for the bottom raw.
As seen from the position of wall, model~1 predicts milder acceleration due to the gamma factor.
Since the wall is accelerated from $v_w = 0.3$, the solution starts from the deflagration regime, goes through the hybrid, and finally forms a detonation profile.

Interestingly, we can clearly see the appearance of the sub-peak except for the top left one, although it eventually decays. 
By referring to Fig.~7 in Ref.~\cite{Espinosa:2010hh}, we can understand that this structure appears when the bubble wall goes beyond ``sound barrier''.
If $v_w$ is increased while $\alpha$ is fixed, the maximum fluid velocity $v_{\rm max}$ in the plasma rest frame changes discontinuously when transitioning from the hybrid region to the detonation region. 
This jump becomes smaller for $\alpha \gtrsim 1$ and therefore the sub-peak does not appear evidently in the top left where the jump takes place $\alpha(t) \sim 1$ due to the moderate acceleration.
To the best of our knowledge, this is the first time that the appearance and characterisation of sub-peaks has been discussed.
Such a sub-structure should affect the spectrum of SGWB from sound waves; see discussions in Sec.~\ref{sec:dc}.

\section{Comparison with the literature
}
\setcounter{equation}{0}
\label{sec:comparison}

\subsection{Evolution in the FLRW background
}
\label{sec:comparison_probe}

In Ref.~\cite{Cai:2018teh}, the profile and the energy budget of the fluid around an expanding bubble are investigated in the FLRW background 
\begin{align}
ds^2
&=
a(\eta)^2(- d\eta^2 + dr^2 + r^2d\Omega^2).
\end{align}
Introducing the self-similar variable $\xi \equiv \eta/r$, the equation
for the fluid velocity $v(\eta, r) = v(\xi)$ is found to be
\begin{align}
2 \frac{v}{\xi} + n \left( 3 + \frac{\mu v}{c_s^2} \right)
&=
\gamma^2 (1 - \xi v) \left( \frac{\mu^2}{c_s^2} - 1 \right) \partial_{\xi} v,
\end{align}
where $n$ is defined as $n \equiv (1 + \eta_0/\eta)^{-1}$ with $\eta_0$ being the time of nucleation.
Note that $n$ explicitly depends on time and therefore the system in general cannot be described in a self-similar manner.
Focusing on the late-time limit of the system and taking $n = 1$ (i.e. $\eta \gg \eta_0$), a self-similar solution from the above EoM is derived, with the junction condition which has the same form as that in the flat spacetime. 
As a result, it is found that the late time solution in FLRW background develops a thinner fluid shell in the self-similar coordinate.

Although their definition of fluid velocity $v(\eta,r)$ is the same as ours, it is not so straightforward to compare the result of Ref.~\cite{Cai:2018teh} to ours because of the following reasons.
First, in our study the condition that the vacuum energy be subdominant requires that $\eta_0/\eta$ to be not too small.
This means that our numerical solution falls in the intermediate regime of $n$ that cannot be handled within the framework of in Ref.~\cite{Cai:2018teh}.
Second, while Ref.~\cite{Cai:2018teh} shows the phase diagram (or the flows) of late-time self-similar solutions, it is not obvious how the actual fluid dynamics traces those flows as the $\alpha(t)$ changes in time.
The formation of shorter tails and the non-convergent behavior to $\xi = c_s$ in their late-time solution for the outward velocity (see Fig.~5 in Ref.~\cite{Cai:2018teh}), in contrast to ours, might come from the difference in the formulation.
Thirdly, our formulation includes the backreaction from the fluid to the spacetime.
Regardless of these differences, the fluid in our intermediate-region simulation ($t \lesssim \mathcal{O}(H_0^{-1}$)) also forms a thinner shell, and this is qualitatively in good agreement with their results.
They also found reduction in the kinetic energy fraction due to the thinning of the profile similar to what we find in this paper.

\subsection{Self-similar solution in GR}
\label{sec:comparison_GR}

In Ref.~\cite{Giombi:2023jqq}, the behavior of two different types of fluid partitioned by an expanding bubble is investigated with the Misner-Sharp formalism, under a spherically symmetric metric
\begin{align}
ds^2
&= - a(t,r) dt^2 + b(t,r)dr^2 + R(t,r)^2d\Omega^2.
\end{align}
Full-GR equations of motion are solved with respect to the variable $\xi \equiv R(t,r)/t$ assuming self-similarity, which significantly simplifies the set of Einstein equations.
After carefully examining the notion of observer and relative velocity, the fluid profile is demonstrated in Figs.~4--9 of Ref.~\cite{Giombi:2023jqq}.
In this study, linear barotropic equations of state are assumed both in the broken phase and symmetric phase ($p = \omega_+\rho$ and $p = \omega_-\rho$, respectively), which can be qualitatively different from what is expected in realistic FOPTs.
These equations of state are in some sense designed to realize a self-similar profile, while the system we analyze has a bag equation of state, the one typically assumed in the study of cosmological FOPTs.
Although the formalism adopted in the present paper cannot be applied to the strong gravity regime as Ref.~\cite{Giombi:2023jqq}, we thus believe that our approach has certain advantages in investigating the dynamics of cosmological FOPTs.

Among the several quantities plotted, velocity $u$ in Ref.~\cite{Giombi:2023jqq} is somewhat similar to the velocity $v$ in the present paper.
The former is defined as the velocity relative to constant-$\xi$ observers defined with respect to the cosmic expansion caused by the fluid outside.
Given the difference in their system and ours mentioned above, their $\xi$ is qualitatively different from our normalized coordinate $r/\eta$, and thus it is again not so straightforward to compare both results.
For example, we cannot immediately see whether the thinning of fluid shell takes place in their results.
Also, while in their setup the equations of motion are solved in such a way that $u$ completely vanishes outside the bubble wall, on general grounds nonzero velocity would in reality be expected outside the wall by the emergence of the new phase.
Nevertheless, our results have some similarity with those in Ref.~\cite{Giombi:2023jqq}. 
From Fig.~\ref{fig:FRW_late}, one can see that our late time solutions develop a slight inward velocity inside the bubble wall for the hybrid and detonation cases.
The similar behavior can be seen in Figs.~6--9 of Ref.~\cite{Giombi:2023jqq}.
On the other hand, the same kind of inward velocity is observed also for deflagration in Ref.~\cite{Giombi:2023jqq} but not in our case.
This could be because the increase in $\alpha(t)$, which is the main reason for the formation of the outward velocity inside the bubble in our study, is absent in their study.

\section{Discussion and conclusions}
\setcounter{equation}{0}
\label{sec:dc}

In this paper, we have investigated general relativistic effects on the sound waves during cosmological FOPTs.
Combining the Higgsless scheme~\cite{Jinno:2020eqg,Jinno:2022mie} with the cosmological hydrodynamics~\cite{Noh:2018sil}, we solve the fluid dynamics including both the cosmological expansion and the perturbative backreaction of fluid to the spacetime, without assuming the self-similarity.
The former method properly integrates out microscopic scales from the system and thus it is easy to combine with hydrodynamic schemes for macroscopic fluid.
We found that, while the gravitational potential that causes slight inward velocity outside the wall, the time-dependence of $\alpha(t)$ driven by the cosmological expansion plays a dominant role. 

We discuss cases with a constant wall velocity in Sec.~\ref{sec:const}.
We find that fluid generally develops a thinner shell in the co-moving coordinate regardless of the wall velocity and the strength of the transition.
A similar prediction was made in Ref.~\cite{Cai:2018teh} by solving the fluid dynamics in the FLRW spacetime background, but the analysis was limited only in the late-time regime.
Our numerical results indicate that this is the case also for an intermediate regime where the bubbles are still of sub-Hubble scales.
To quantify this thinning, we scanned the input parameters $(v_w,\alpha_0)$ to evaluate the kinetic energy fraction $\kappa(t)$ at each time slice as summarized in Figs.~\ref{fig:contour1}--\ref{fig:contour3}.
These figures illustrate how the energy budget for the fluid kinetic energy could be reduced over the cosmological time scales.

Besides the thinning effect on the profile, we observe other kinds of interesting behavior by virtue of not assuming self-similarity in the simulations.
In the case of deflagrations, we find the formation of tails in the fluid velocity behind the wall is characteristic behavior.
We also find that profiles originally in the detonation regime can eventually develop into the hybrid regime.
Both of them are mainly driven by the evolution of $\alpha(t)$.
To investigate more realistic situations, we also solved the system with time-dependent wall velocity $v_w(t)$ in Sec.~\ref{sec:wall_acc}.
Interestingly, we find that the fluid can develop a spiky substructure as the profile changes from the hybrid to the detonation regime.
To the best of our knowledge, the appearance of this spike has been reported for the first time.

Our results have several implications to the SGWB production.
From calculations in the flat background, it is known that the GW spectrum develops a plateau between (the inverse of) the typical bubble size to (the inverse of) the fluid shell thickness~\cite{Hindmarsh:2013xza,Hindmarsh:2015qta,Hindmarsh:2017gnf,Cutting:2019zws,Jinno:2020eqg,Jinno:2022mie}\cite{Hindmarsh:2016lnk,Hindmarsh:2019phv,RoperPol:2023dzg}.
Since the GW production in the FOPT tends to be dominated by largest bubbles, and since the fluid shell becomes thinner for such bubbles as found in the present study, the GW spectrum for FOPT scenarios with large bubbles may have a broader plateau than previously thought.
Also, the reduction of kinetic energy fraction caused by the thinning of the fluid shell may result in a decrease of the SGWB amplitude.
Moreover, the peaky substructure that appears when the accelerating wall goes beyond the sound barrier could lead to more complicated structures in the GW spectrum.
While in this study we consider the profile of a single bubble expanding without collision, actual FOPTs end with collision with other bubbles.
Gravitational effects on the bubbles are thus cut off by the timescale of the transition, or equivalently the typical bubble size, determined by the parameter $\beta / H = d(S_3/T) / d \ln T$ and higher derivatives $d^n(S_3/T) / d (\ln T)^n$.
Ultimately, all these aspects must be taken into account in the GW templates when reconstructing the fundamental model parameters from the GW signals once observed.
We leave detailed study on this point for future work.

\section*{Acknowledgments}

The authors thank Naoki Yoshida, Thomas Konstandin, Isak Stomberg, Lorenzo Giombi, Mark Hindmarsh, and Henrique Rubira for fruitful discussions.
The work of RJ is supported by JSPS KAKENHI Grant Numbers 23K17687, 23K19048, and 24K07013.
JK is supported by the JSPS Overseas Research Fellowships.
JK acknowledges support from Istituto Nazionale di Fisica Nucleare (INFN) through the Theoretical Astroparticle Physics (TAsP) project.
\appendix

\section{Comparison with a commonly used gauge choice}
\label{sec:gauge_choice}
In the cosmological perturbation theory, one common choice is the zero-shear gauge (ZSG), in which the temporal gauge condition is chosen such that $\chi = 0$ instead of $\kappa = 0$.
In this gauge, the variables $\Phi$ and $\Psi$ correspond to the well-known Bardeen potentials~\cite{Bardeen:1980kt}.
While the gauge-invariant scalar potential of velocity perturbation $v_0(t,r)$ (satisfying $v(t,r) = \partial_r v_0(t,r)$) is identical in both ZSG and uniform-expansion gauge (UEG), the other quantities differ as
\begin{equation}
\Phi_{\rm ZSG} = \Phi + \dot{\chi},
\qquad
\Psi_{\rm ZSG} = \Psi + H \chi,
\qquad
\delta w_{\rm ZSG} = \delta w - 4 a H w_b \left( v_0 + \frac{\chi}{a} \right),
\label{eq:gauge_inv}
\end{equation}
where the quantities in the right-hand side are measured in the UEG. 
However, we argue that their difference is small in the following sense.
On general ground we expect
\begin{equation}
\Phi
\sim
\Psi
\sim
\frac{\delta w}{w_b}
\sim
v.
\end{equation}
On the other hand, as discussed in Eq.~(\ref{eq:chi_perturbative}), we
require the condition
\begin{equation}
\frac{\partial_r \chi}{a}
\sim
\frac{\chi}{l_c}
\ll
v,
\end{equation}
with $l_c$ being the characteristic length scale in the present system. In the regime of our interest, we can identify it with the sound shell thickness.
Therefore, the differences in the two gauges behave as
\begin{equation}
\dot{\chi}
\sim
H \chi
\ll
H l_c v,
\qquad
a H w_b \left( v_0 + \frac{\chi}{a} \right)
\sim
a H w_b \left( \frac{l_c v}{a} + \frac{\chi}{a} \right)
\sim
H l_c w_b v,
\end{equation}
and thus we expect a suppression factor of $H l_c < 1$.
We finally note that as discussed in Refs.~\cite{Hwang:2016kww,Noh:2018sil}, zero-shear gauge results in the inconsistency with the spherical symmetric solution of Tolman-Oppenheimer-Volkoff~\cite{PhysRev.55.364, PhysRev.55.374} when the fluid has relativistic pressure.
Notice that this issue does not exist in the uniform-expansion gauge, which provides additional motivation for adopting it in our system.

\section{True vacuum region as the background}
\label{sec:tv_as_bg}

To compare the result in Sec.~\ref{sec:with_GR}, here we assume the background consists only of radiation as 
\begin{align}
p_b
&=
\frac{1}{3}\rho_b
=
\frac{1}{4}w_b, 
\end{align}
which yields the background dynamics $\rho_b \propto 1/a^4(t)$ and $ H \propto a(t)^{-2}$.
In this case, the vacuum energy $\epsilon$ is now included in $\delta\rho$ and $\delta p$ and we can assume the relation 
\begin{align}
\delta \tilde{p}
&=
\frac{1}{4}\delta \tilde{w} - \frac{3}{4}\alpha(t) \Theta(r - r_w(t)),
\\
\delta \tilde{\rho}
&=
\frac{3}{4}\delta \tilde{w} + \frac{3}{4}\alpha(t) \Theta(r - r_w(t)).
\end{align}
Note that the fundamental variables $(\tilde{T}^t, \tilde{T}^r)$ are defined in the same way as in Sec.~\ref{sec:cosmo_hydro_FOPT} and its relation to the other quantities is also preserved.
The only difference arises from the expression of $\delta\tilde{p}$, which affects the gravitational potential $\Phi$ via Eq.~\eqref{eq:eom_grav_pot}.

\begin{figure}[htbp]
\begin{center}
\includegraphics[width=0.8\columnwidth]{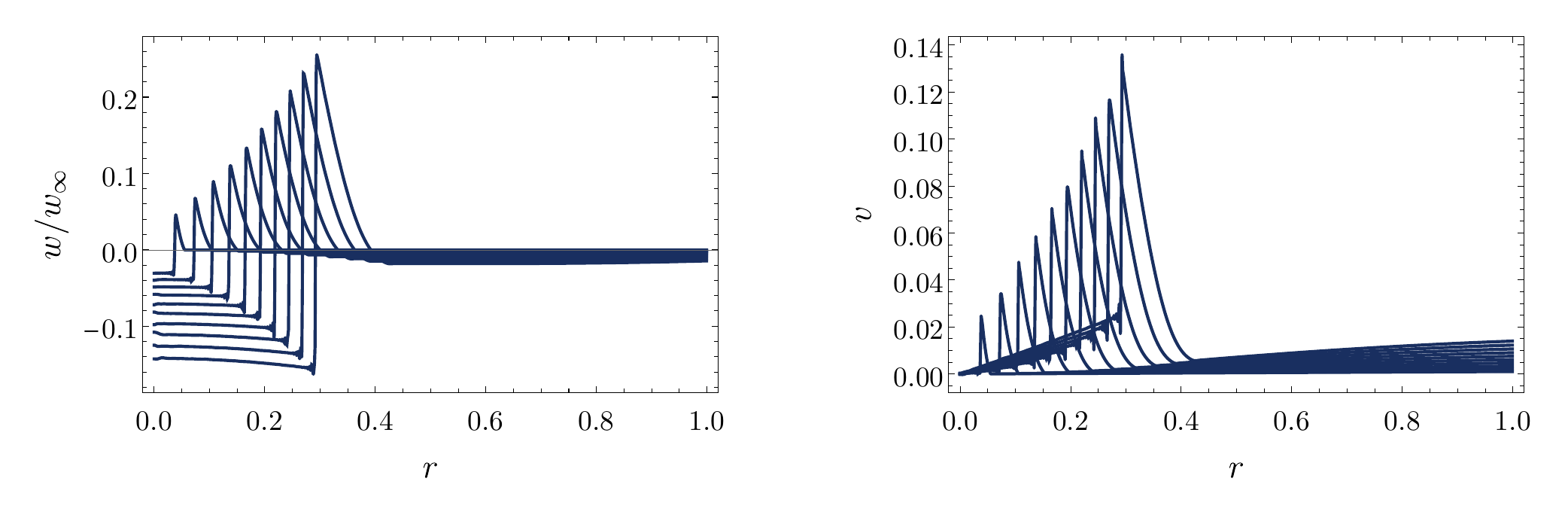}
\includegraphics[width=0.8\columnwidth]{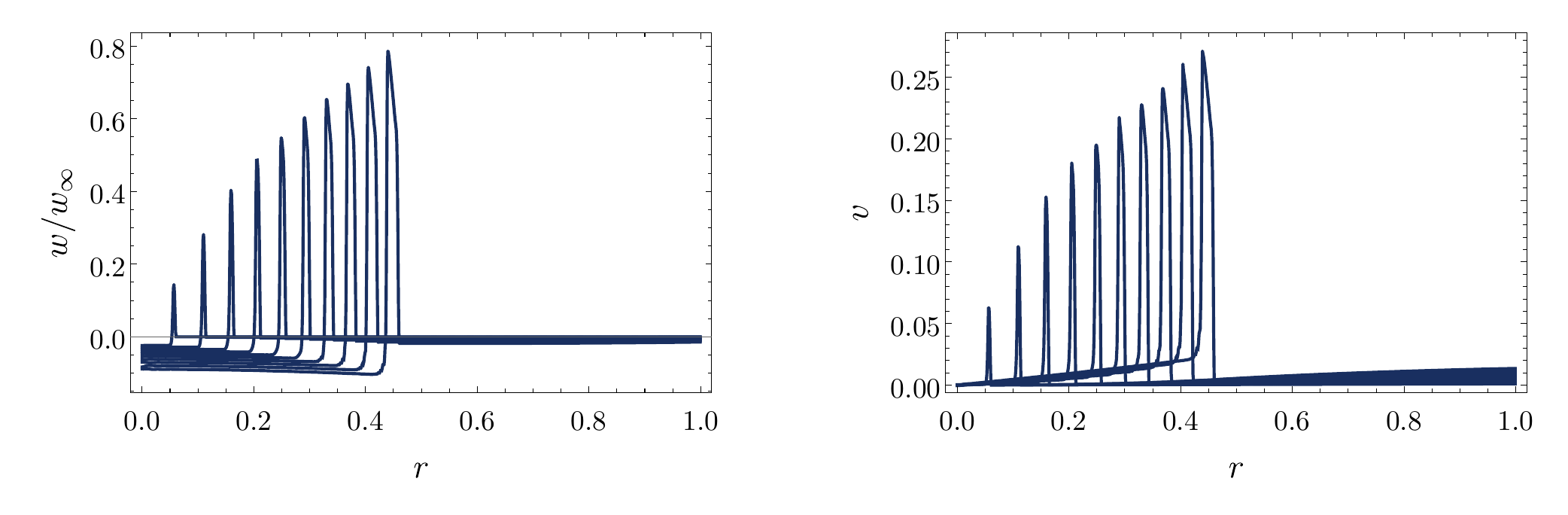}
\includegraphics[width=0.8\columnwidth]{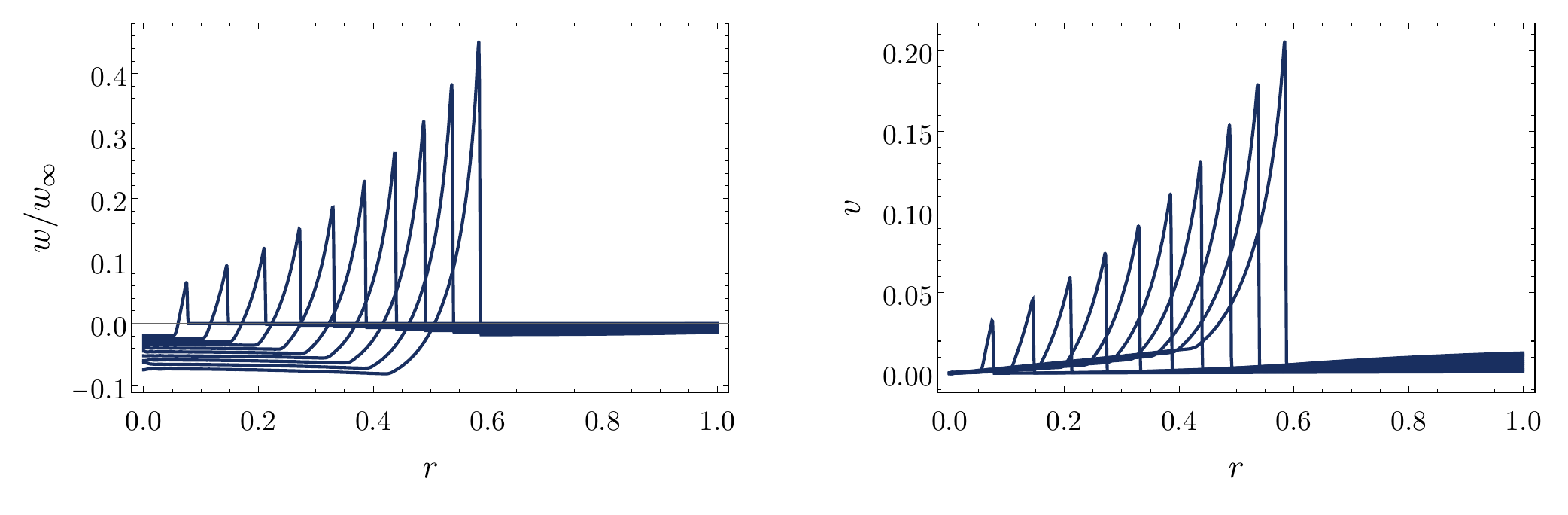}
\caption{\small
Fluid profile simulated at the parameter points similar to those in Fig.~\ref{fig:FRW2}.
}
\label{fig:FRW}
\end{center}
\end{figure}

In the numerical calculation, we assume the evolution of scale factor as $a(t) = \sqrt{1 + t/t_0}$. 
A parameter $t_0$ characterizes the expansion rate and the background enthalpy as $H(t) = 1/2(t + t_0) = (1/2t_0)a(t)^{-2}$ and $w_b/M_{\rm Pl}^2 = 1/(t + t_0)^2 = (1/t_0^2)a(t)^{-4}$ respectively.
In the present case, we find the analytical expression of wall position~\eqref{eq:wall_position} as
\begin{align}
r_w(t)
&=
2 t_0^{1/2} v_w \left[ (t + t_0)^{1/2} - t_0^{1/2} \right].
\end{align}
Note that $t_0$ is related to the initial Hubble constant as $H_0 = 1/2t_0$.
We show in Fig.~\ref{fig:FRW} the time evolution of a spherical expanding bubble simulated at the same parameter points as Fig.~\ref{fig:FRW2}, namely $\alpha_0 = 10^{-2}$ and $v_w = (0.4, 0.6, 0.8)$. 

As in Fig.~\ref{fig:FRW2}, we find the non self-similar evolution of the fluid and the formation of the tail in the velocity inside the wall, which are due to the time-dependence in $\alpha(t)$.
Outside the walls, however, the behaviour seems totally different.
One can see the the existence of outward velocity outside the wall and the corresponding decrease in the enthalpy.
These can be explained by the accelerating expansion caused by the vacuum energy. 
Since we take only radiation as the background, the outward velocity is sourced by the gravitational potential $\Phi$ as the deviation from the actual cosmological expansion. 

\begin{figure}[htbp]
\begin{center}
\includegraphics[width=0.5\columnwidth]{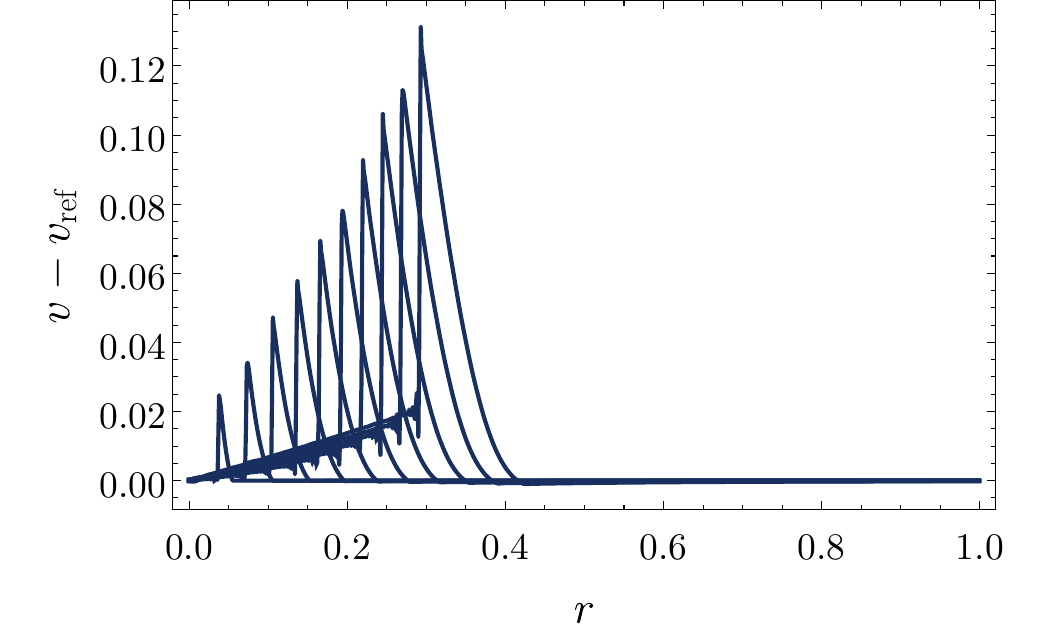}~
\includegraphics[width=0.5\columnwidth]{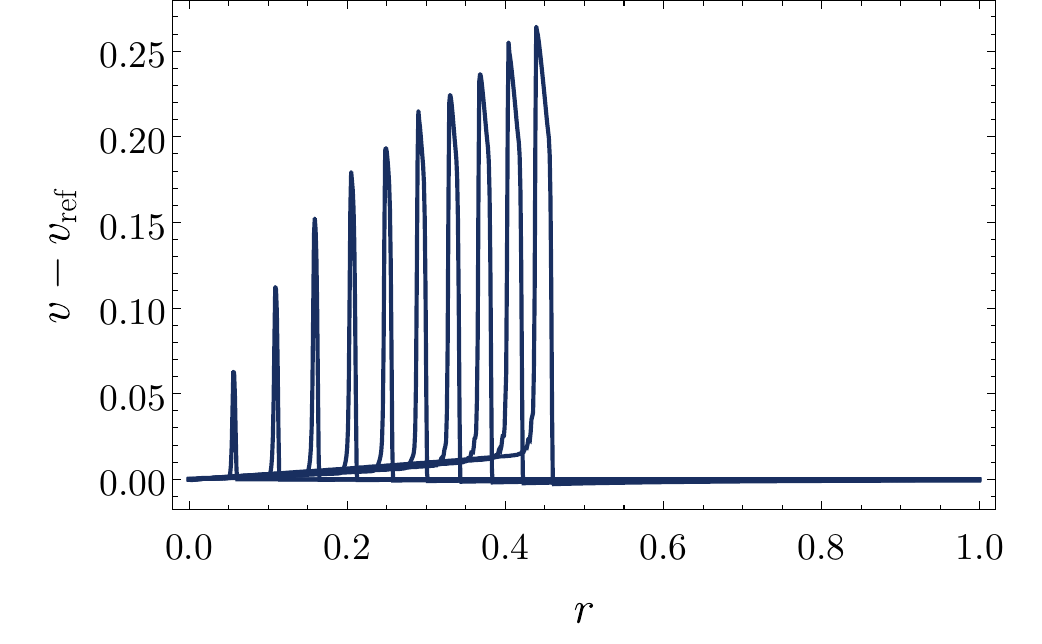}\\
\includegraphics[width=0.5\columnwidth]{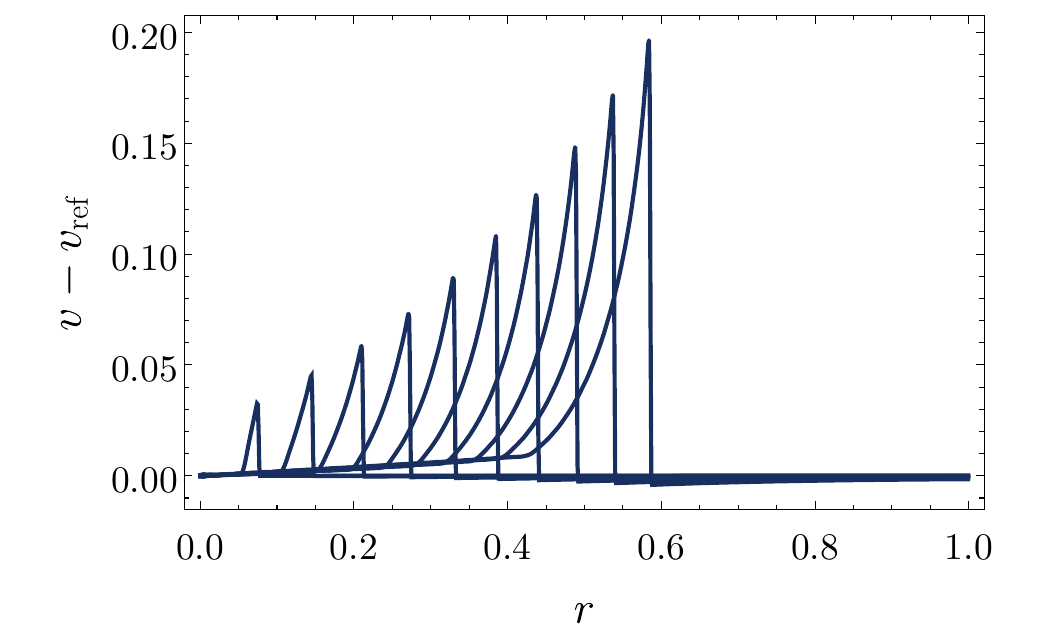}
\caption{\small
After the subtraction of the reference velocity $v_{\rm ref}$. If linearity holds well, the velocity profile after this operation should correspond to the peculiar motion of the fluid. Indeed, it can be seen that this matches well with the right panels of Fig.~\ref{fig:FRW2}.
}
\label{fig:FRW_ref_sub}
\end{center}
\end{figure}
Note that such a global acceleration cannot be treated as a peculiar motion and might be incompatible with the short-scale assumption in Eq.~\eqref{eq:conditions}.
Indeed, the slight decrease in the enthalpy outside the wall, which was not observed in Fig.~\ref{fig:FRW2}, is caused by this outward velocity and one should be a little careful when interpreting this result.
Nevertheless, as shown in Fig.~\ref{fig:FRW_ref_sub}, we find the result of velocity profile becomes quite consistent with those in Fig.~\ref{fig:FRW2}, after a simple operation.
That is, we simply subtract the reference velocity $v_{\rm ref}(t,r)$, which can be obtained by solving the system with $r_w(t) = 0$, from the velocity profiles in Fig.~\ref{fig:FRW}.
We believe that with this agreement, our result on the fluid peculiar velocity in Sec.~\ref{sec:with_GR} becomes more convincing.

\section{Late time evolution}
\label{app:late_time}

\begin{figure}[htbp]
\begin{center}
\includegraphics[width=0.45\columnwidth]{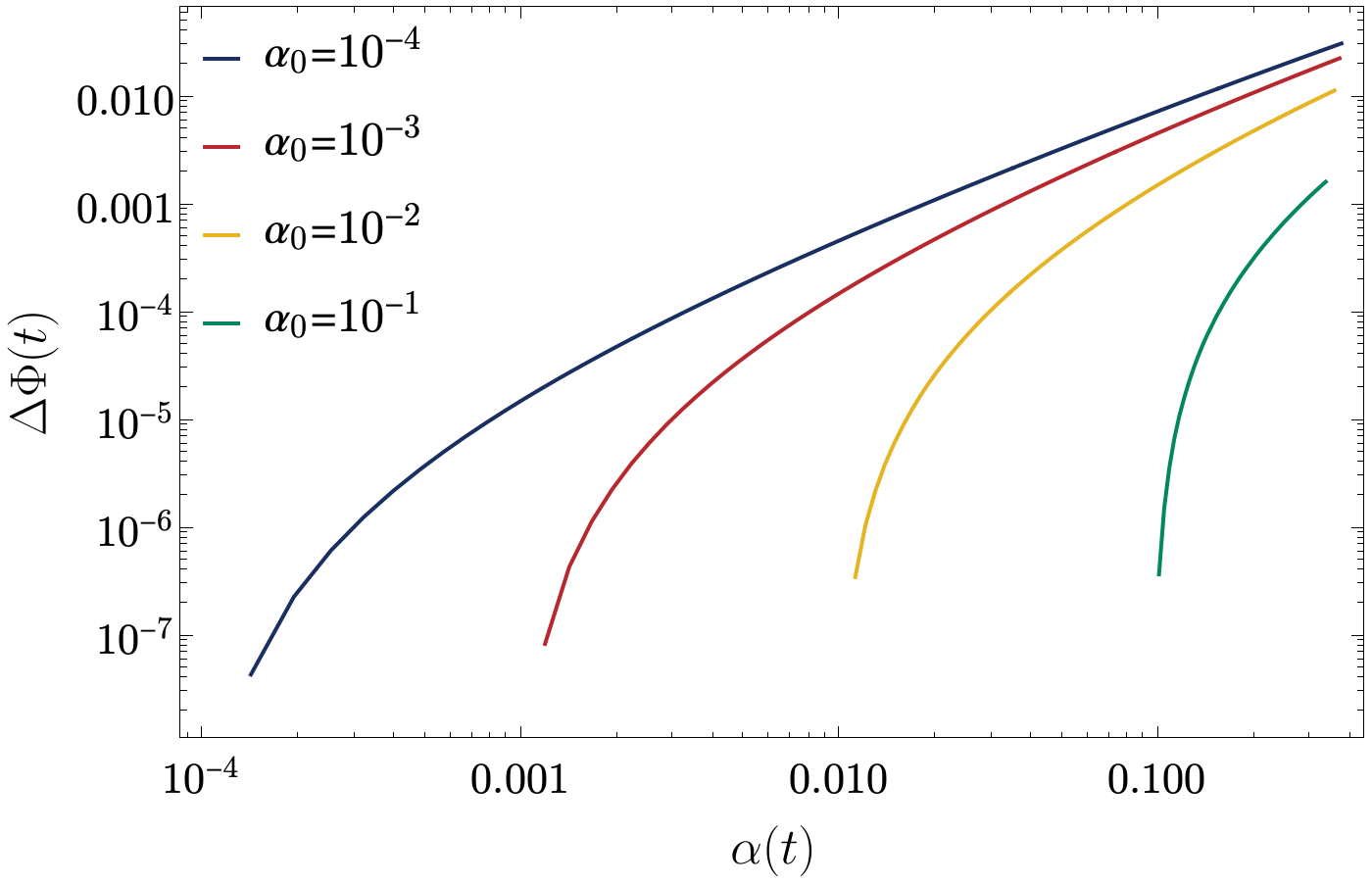}
\includegraphics[width=0.45\columnwidth]{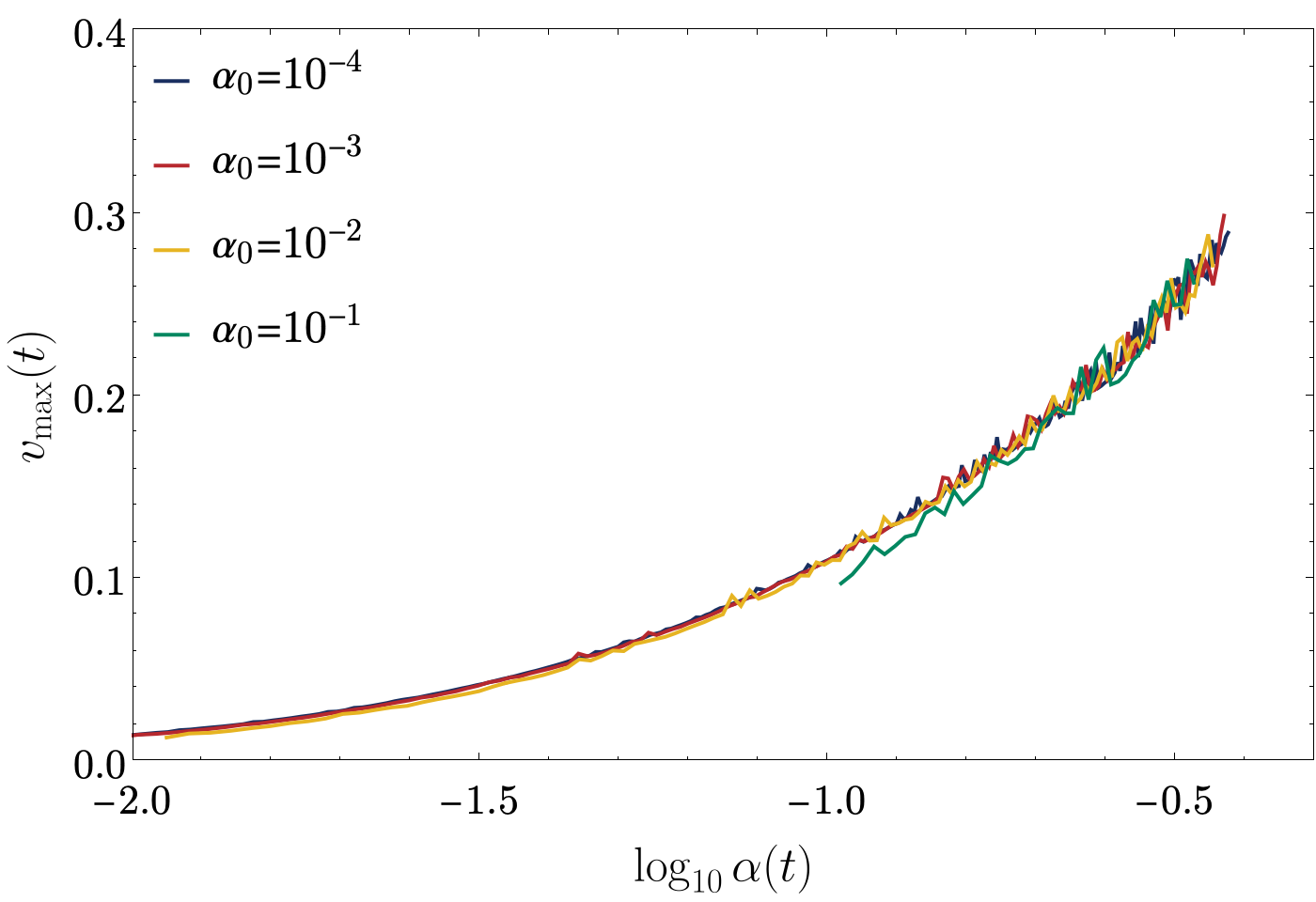}\\
\includegraphics[width=0.45\columnwidth]{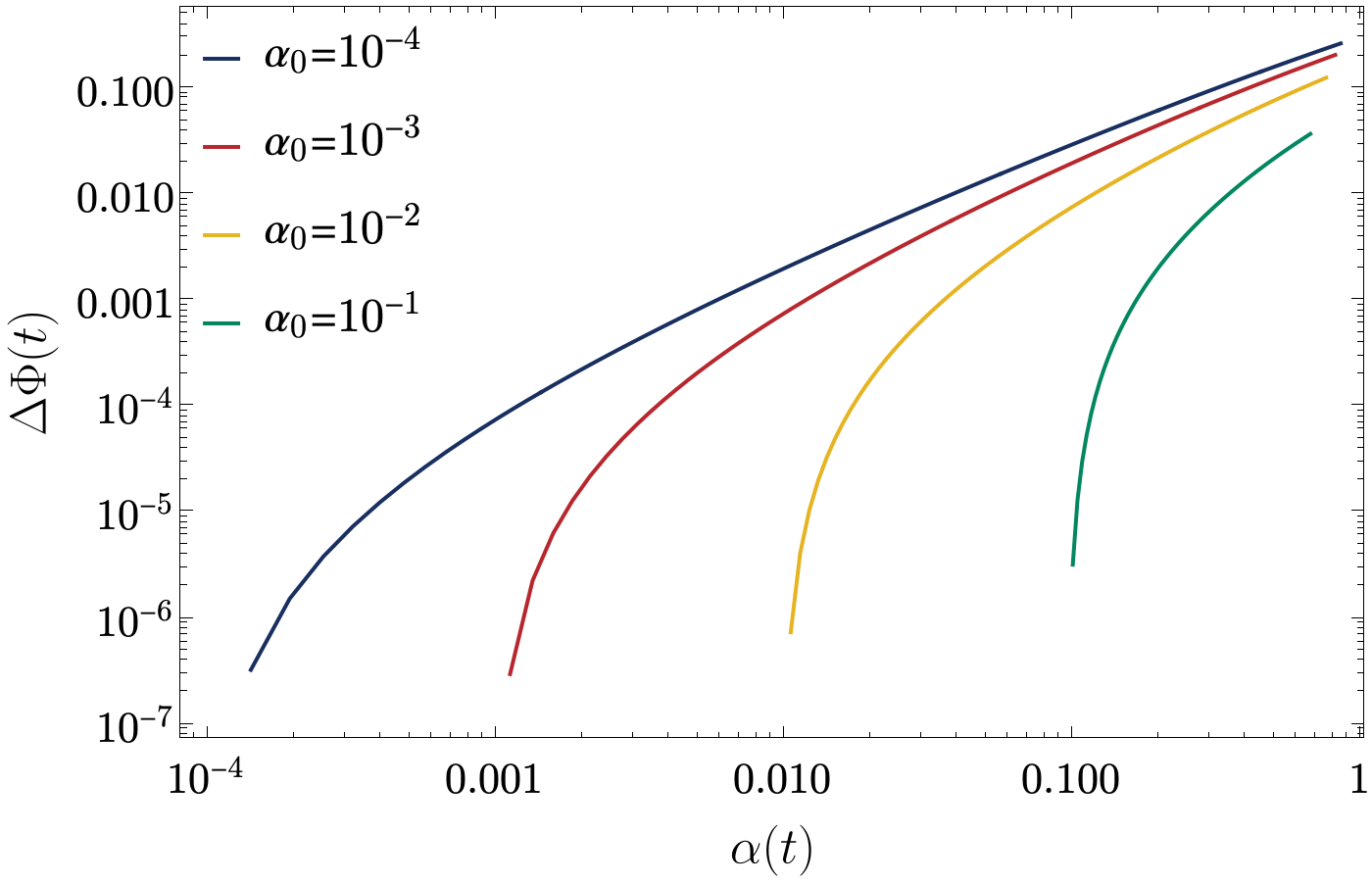}
\includegraphics[width=0.45\columnwidth]{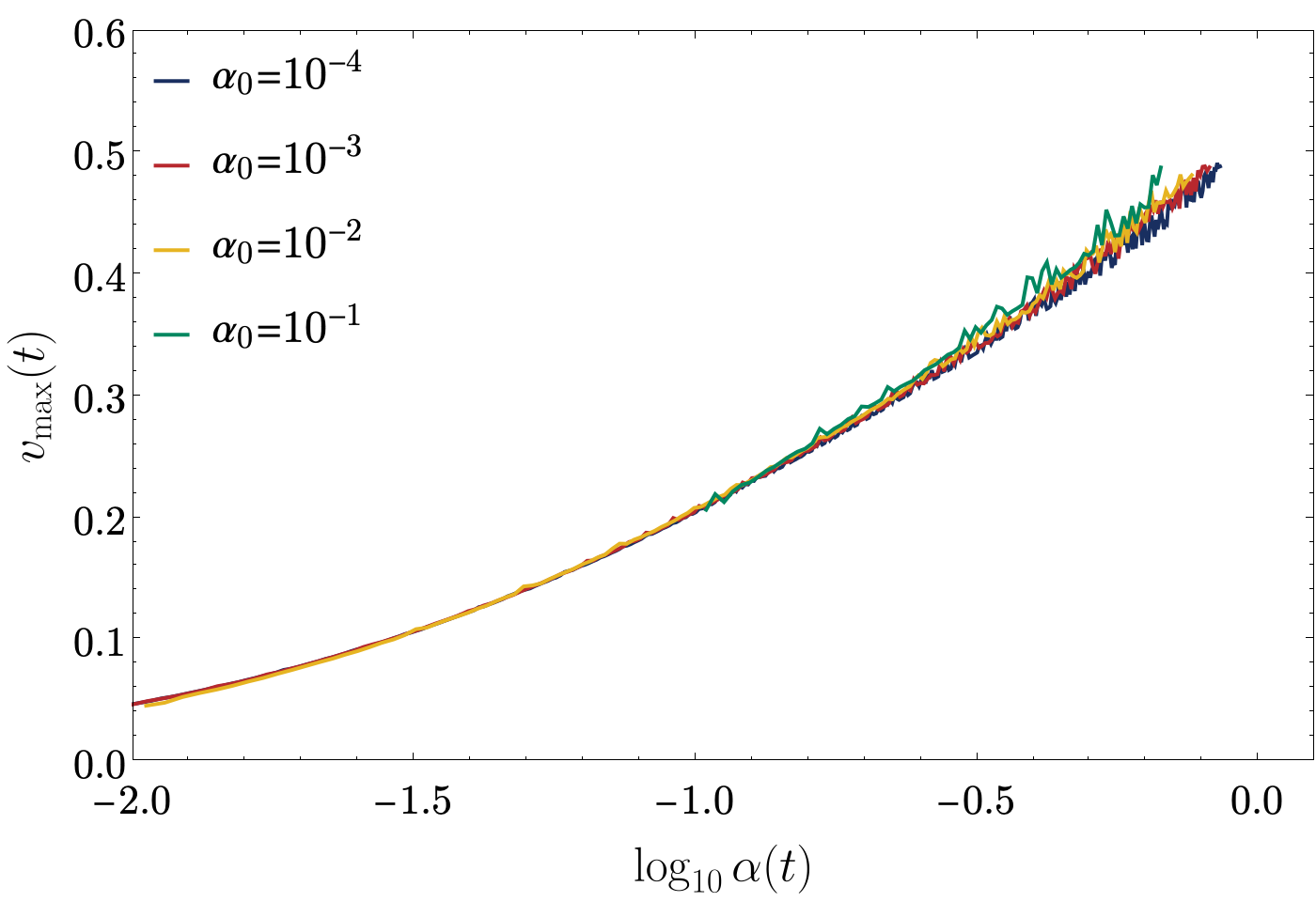}\\
\includegraphics[width=0.45\columnwidth]{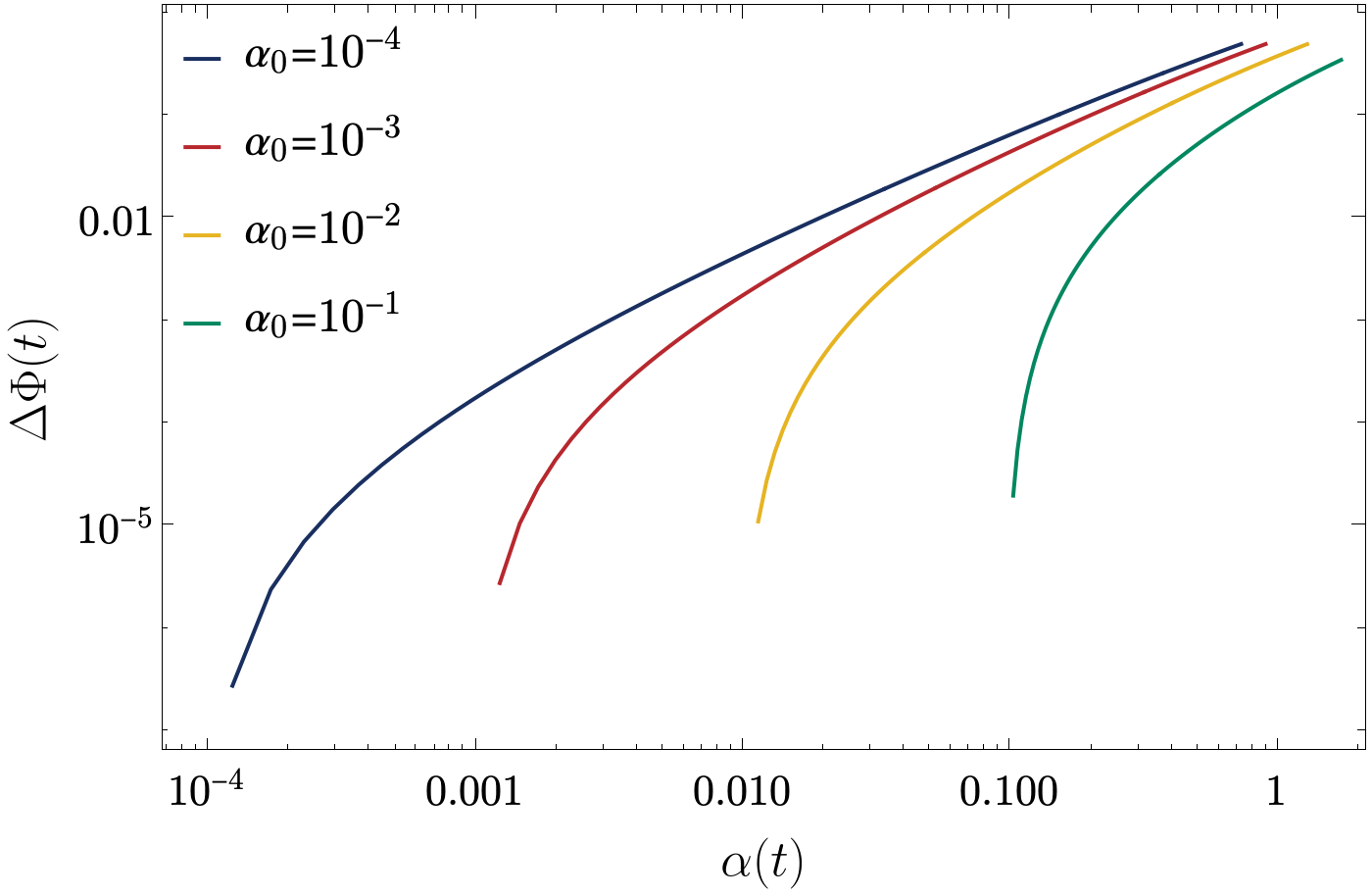}
\includegraphics[width=0.45\columnwidth]{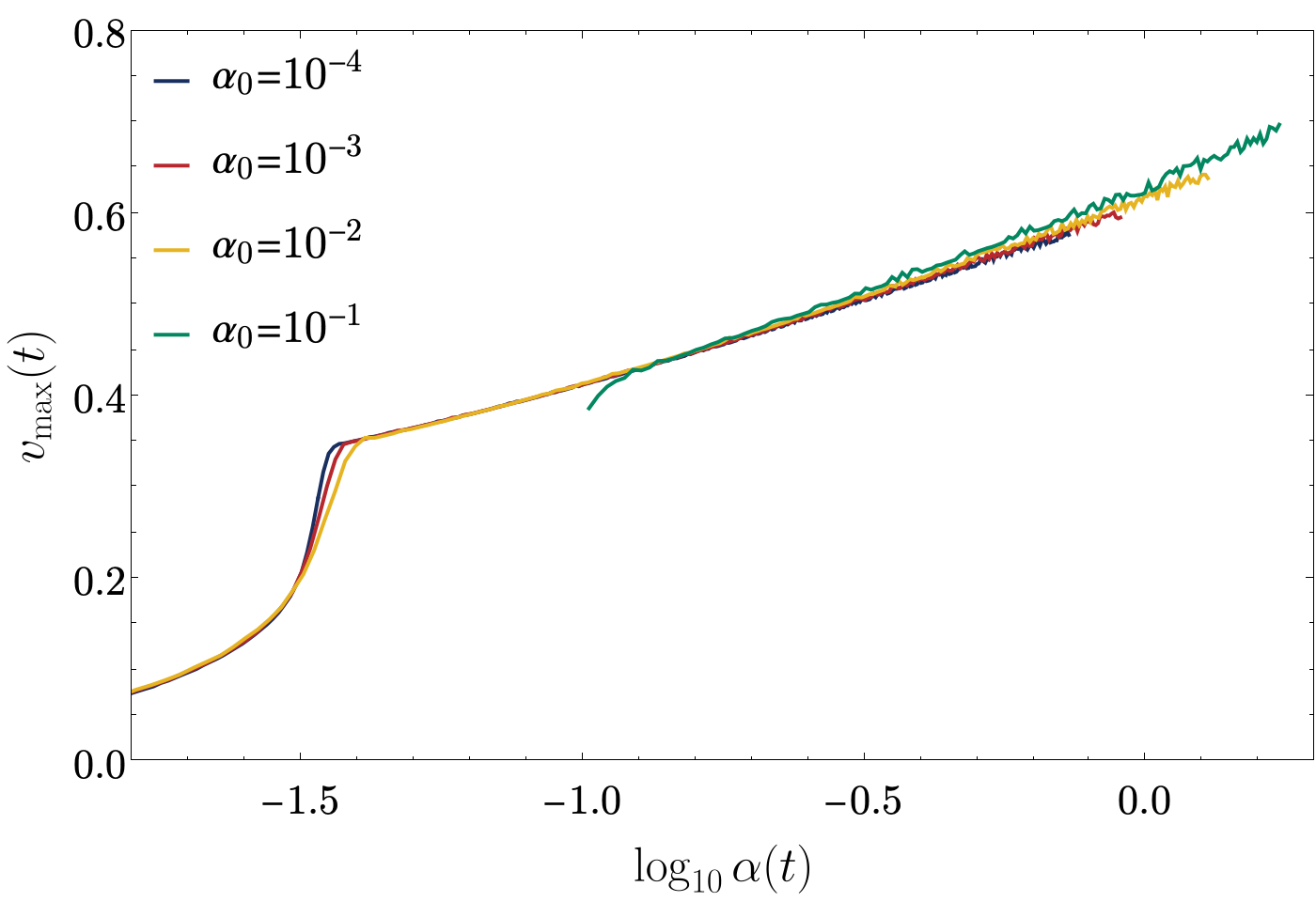}\\
\includegraphics[width=0.45\columnwidth]{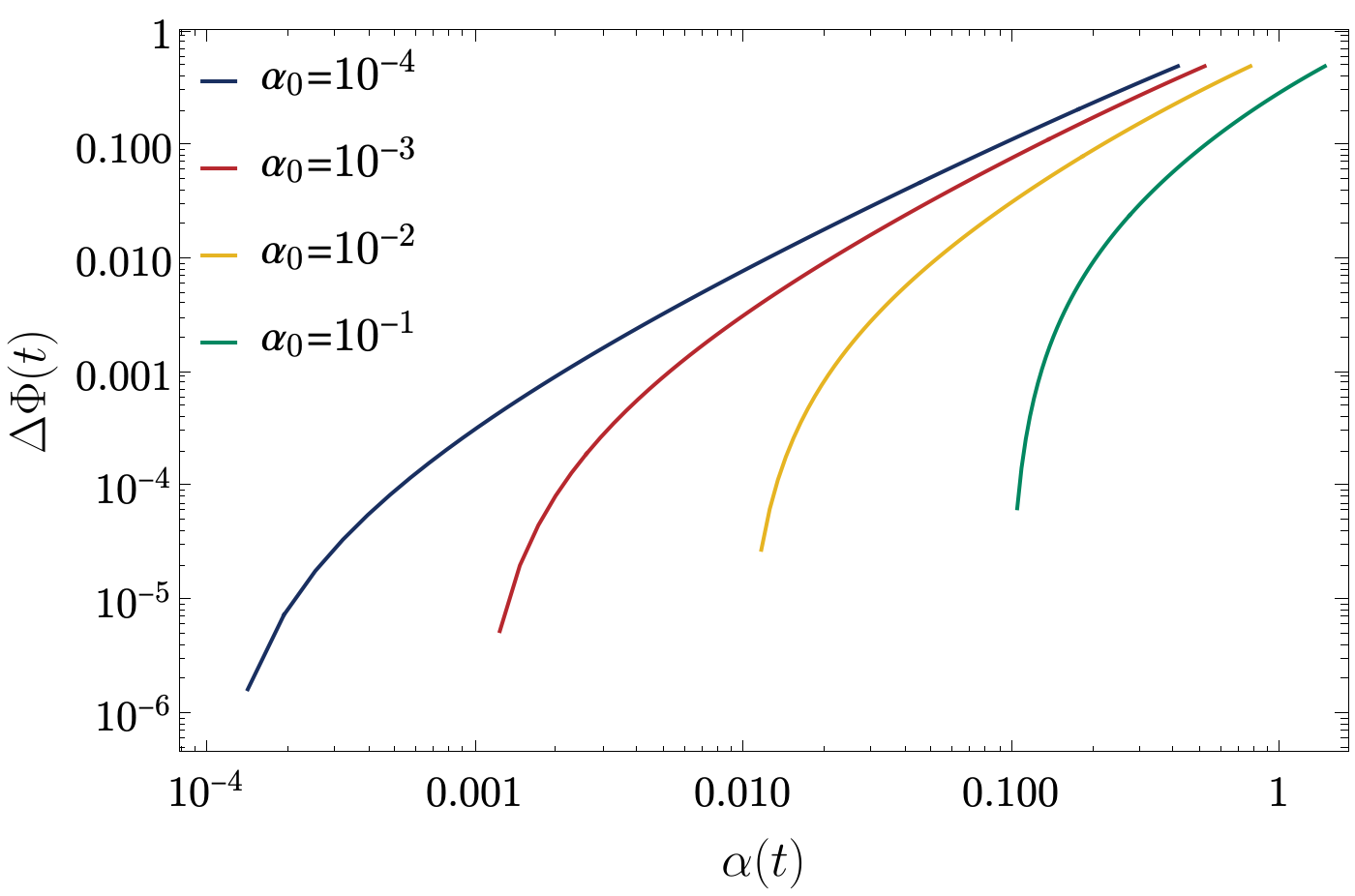}
\includegraphics[width=0.45\columnwidth]{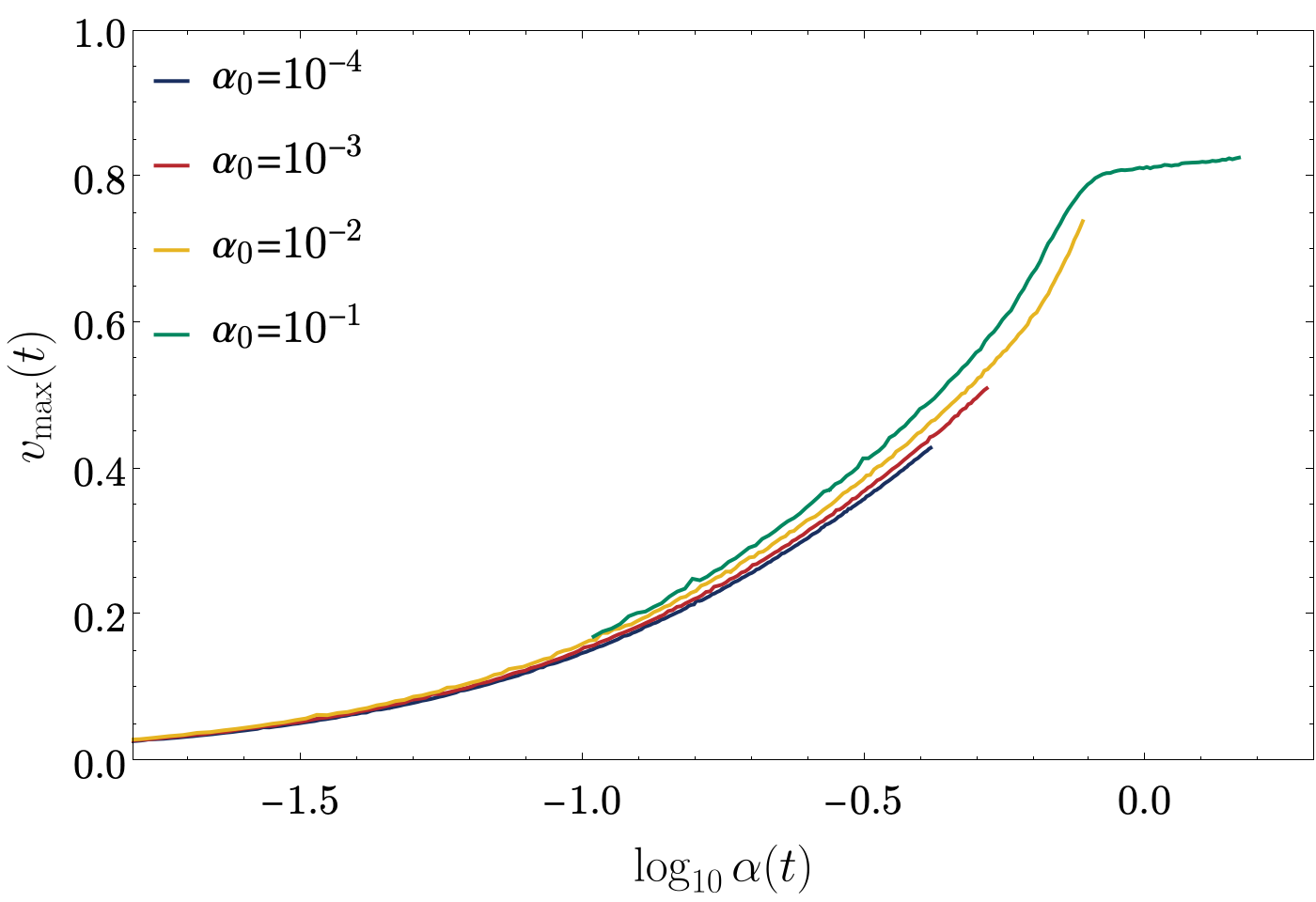}
\caption{\small
The evolution of $\Delta\Phi$ (left panel), $v_{\rm max}$ (right panel) against the different initial condition $\alpha_0$ are shown for the wall velocity $v_w = 0.5, 0.7, 0.9$ (from top to bottom raw). 
}
\label{fig:luminal}
\end{center}
\end{figure}

In Fig.~\ref{fig:luminal}, we show the evolution of maximum value of gravitational potential $\Delta \Phi(t)$ 
and maximum fluid velocity $v_{\rm max}(t)$ over the cosmological time scales with different input parameters.
As discussed in the main text, we observed the convergent behavior of the late time value of $\kappa(t)$ when taking smaller value of $\alpha_0$.
Here we can see the corresponding behavior in these quantities.
That is, late time trajectories become closer for smaller $\alpha_0$, indicating that the late-time behavior becomes less sensitive to when precisely the bubble nucleates.
Note that we observe the zigzags in the late time $v_{\rm max}(t)$ for $v_w = 0.5\ \&\ 0.7$. This shows that the performance of Kurganov-Tadmor shock capturing degrades at larger $\alpha$ for deflagration and hybrid profiles. 
However, this error does not much affect the estimate of $\kappa(t)$.

For $v_w = 0.7$ and $v_w = 0.9$, where the simulations are terminated at $\Delta \Phi(t) = 0.5$, we can clearly see the horizontal shift of the end points of the simulation in the trajectory of $\Delta \Phi(t)$. 
This is a reasonable behavior since $r_w(t)$ at some fixed value of $\alpha(t)$ becomes larger for smaller $\alpha_0$, resulting in the faster growth of $\Phi(t)$. Notice that the amount of shift decreases when $\alpha_0$ becomes smaller.
For $v_{\rm max}$, we can clearly see a transition at $\alpha(t) \sim 10^{-1.5}$ and $\alpha(t) \sim 10^{-0.1}$ respectively for $v_w = 0.7$ and $v_w = 0.9$.
These are the points where the fluid profile transitions from the detonation to the hybrid. We also see the slight decrease in the value of $v_{\rm max}$ when decreasing $\alpha_0$, which is caused by the large gravitational potential $\Phi(t)$. The rate of decrease again becomes smaller for smaller $\alpha_0$ showing the convergent behavior.

For $v_w = 0.5$ (and in fact $v_w \lesssim c_s$), the simulations are terminated by the condition $v_{\rm max} < v_w$ before $\Delta \Phi(t)$ reaching 0.5. 
Therefore, we observe a vertical shift in the end points of the simulation in the trajectory of $\Delta \Phi(t)$.
We again observe the decreasing behavior in the late time $v_{\rm max}$ caused by large $\Phi$.

\small
\bibliographystyle{JHEP}
\bibliography{ref}

\end{document}